\documentclass[preprint,aps,showpacs,floatfix,superscriptaddress,nofootinbib]{revtex4}
\usepackage[dvips]{graphicx}
\usepackage{epsf}
\usepackage{enumerate}
\usepackage{amsmath}
\usepackage{amssymb}
\usepackage{amsfonts}
\usepackage{latexsym}
\usepackage{revsymb}
\usepackage{bm}
\usepackage{multirow}
\usepackage{axodraw}
\newcommand{\bra}[1]{\left\langle #1 \right|}
\newcommand{\ket}[1]{\left| #1 \right\rangle}

\newcommand {\bea} {\begin{eqnarray}}
\newcommand {\eea} {\end{eqnarray}}
\newcommand {\ba} {\begin{array}}
\newcommand {\ea} {\end{array}}

\def\vsk#1{\noalign{\vskip#1 cm}}

\begin{document}

\preprint{OCHA--PP--273}
\preprint{KEK--TH--1149}
\preprint{VPI--IPNAS--07--06}

\title{Constraints on New Physics from\\Long Baseline Neutrino Oscillation Experiments}
\author{Minako~Honda}\email{minako@hep.phys.ocha.ac.jp}
\affiliation{Physics Department, Ochanomizu University, Tokyo 112-8610, Japan}
\author{Yee~Kao}\email{ykao@vt.edu}
\affiliation{Institute for Particle, Nuclear, and Astronomical Sciences, Physics Department, Virginia Tech, Blacksburg VA 24061, USA}
\author{Naotoshi~Okamura}\email{naotoshi.okamura@kek.jp}
\affiliation{KEK Theory Division, Tsukuba 305-0801, Japan}
\author{Alexey~Pronin}\email{apronin@vt.edu}
\affiliation{Institute for Particle, Nuclear, and Astronomical Sciences, Physics Department, Virginia Tech, Blacksburg VA 24061, USA}
\author{Tatsu~Takeuchi}\email{takeuchi@vt.edu}
\affiliation{Institute for Particle, Nuclear, and Astronomical Sciences, Physics Department, Virginia Tech, Blacksburg VA 24061, USA}

\date{July 31, 2007}

\begin{abstract}
\noindent
New physics beyond the Standard Model
can lead to extra matter effects on neutrino oscillation
if the new interactions distinguish among the three flavors of neutrino.
In a previous paper \cite{HOT}, we argued that a long-baseline neutrino oscillation experiment 
in which the Fermilab-NUMI beam in its high-energy mode \cite{NUMI}
is aimed at the planned Hyper-Kamiokande detector \cite{HyperK}
would be capable of constraining the size of those extra effects, provided the vacuum value of $\sin^2 2\theta_{23}$ is not too close to one.
In this paper, we discuss how such a constraint would translate into limits on the coupling constants and masses of new particles in various models. 
The models we consider are: models with generation distinguishing $Z'$s such as 
topcolor assisted technicolor, 
models containing various types of leptoquarks, 
R-parity violating SUSY, and extended Higgs sector models. 
In several cases, we find that the limits thus obtained could be competitive with those expected from direct searches at the LHC.
In the event that any of the particles discussed here are discovered at the LHC,
then the observation, or non-observation, of their matter effects could help in identifying what type of particle had been observed.
\end{abstract}


\maketitle
\section{Introduction}

When considering matter effects on neutrino oscillation, it is customary to consider only the $W$-exchange interaction of the $\nu_e$ with the electrons in matter.
However, if new interactions beyond the Standard Model (SM) that distinguish among the three generations of neutrinos exist, they can lead to extra matter effects via radiative corrections to the
$Z\nu\nu$ vertex, which effectively violate neutral current universality, 
or via the direct exchange of new particles between the neutrinos and matter particles \cite{Zee:1985rj}.  

Many models of physics beyond the SM introduce interactions which distinguish among generations:
gauged $L_\alpha - L_\beta$ 
\cite{L1minusL2} and gauged $B-\alpha L_e -\beta L_\mu - \gamma L_\tau$ 
\cite{Bminus3Ltau,Bminus3Le,Bminus3over2LtauplusLmu,Chang:2000xy} 
models introduce $Z'$s and Higgs sectors 
which distinguish among the three generations of leptons;
topcolor assisted technicolor treats the third generation differently from the first two to explain the large top mass \cite{TopTechni,TopTechni_limits};
R-parity violating couplings in supersymmetric models couple fermions/sfermions from different generations \cite{Rparity_notations,Barbier:2004ez,Rparity_limits}.  

The effective Hamiltonian that governs neutrino oscillation
in the presence of neutral-current lepton universality violation, or new physics
that couples to the different generations differently, is given by~\cite{HOT}
\begin{equation}
H = 
\tilde{U}
\left[ \begin{array}{ccc} \lambda_1 & 0 & 0 \\
                          0 & \lambda_2 & 0 \\
                          0 & 0 & \lambda_3
       \end{array}
\right]
\tilde{U}^\dagger
= U
\left[ \begin{array}{ccc} 0 & 0 & 0 \\
                          0 & \delta m^2_{21} & 0 \\
                          0 & 0 & \delta m^2_{31}
       \end{array}
\right]
U^\dagger +
\left[ \begin{array}{ccc} a & 0 & 0 \\
                          0 & 0 & 0 \\
                          0 & 0 & 0 
       \end{array}
\right] +
\left[ \begin{array}{ccc} b_e & 0 & 0 \\
                          0 & b_\mu & 0 \\
                          0 & 0 & b_\tau 
       \end{array}
\right] \;.
\label{Hdef}
\end{equation}
In this expression, $U$ is the MNS matrix \cite{MNS},
\begin{equation}
a=2E V_{CC}\;,\qquad V_{CC} = \sqrt{2} G_F N_e = N_e \dfrac{g^2}{4 M_W^2}\;,
\end{equation}
is the usual matter effect due to $W$-exchange
between $\nu_e$ and the electrons \cite{MSW}, and  $b_e$, $b_\mu$, $b_\tau$
are the extra matter effects which we assume to be 
flavor diagonal and non-equal. 
The matter effect terms in this Hamiltonian can always be written as
\begin{eqnarray} \label{matter_hamiltonian}
\lefteqn{\left[ \begin{array}{ccc} a & 0 & 0 \\
                          0 & 0 & 0 \\
                          0 & 0 & 0 
       \end{array}
\right] +
\left[ \begin{array}{ccc} b_e & 0 & 0 \\
                          0 & b_\mu & 0 \\
                          0 & 0 & b_\tau 
       \end{array}
\right]}\cr
& = &
\left[ \begin{array}{ccc} \left(a+b_e-\dfrac{b_\mu+b_\tau}{2}\right) & 0 & 0 \\
                          0 & \left(\dfrac{b_\mu-b_\tau}{2}\right) & 0 \\
                          0 & 0 & -\left(\dfrac{b_\mu-b_\tau}{2}\right) 
       \end{array}
\right] + \left(\dfrac{b_\mu+b_\tau}{2}\right)
\left[ \begin{array}{ccc} 1 & 0 & 0 \\
                          0 & 1 & 0 \\
                          0 & 0 & 1 
       \end{array}
\right] \;.
\end{eqnarray}
The unit matrix term does not contribute to neutrino oscillation so it can be dropped.
We define the parameter $\xi$ as
\begin{equation}
\dfrac{b_\tau - b_\mu}{a} = \xi \;.
\label{xi-def}
\end{equation}
Then, the effective Hamiltonian can be written as
\begin{equation}
H = 
\tilde{U}
\left[ \begin{array}{ccc} \lambda_1 & 0 & 0 \\
                          0 & \lambda_2 & 0 \\
                          0 & 0 & \lambda_3
       \end{array}
\right]
\tilde{U}^\dagger
= U
\left[ \begin{array}{ccc} 0 & 0 & 0 \\
                          0 & \delta m^2_{21} & 0 \\
                          0 & 0 & \delta m^2_{31}
       \end{array}
\right]
U^\dagger + a
\left[ \begin{array}{ccc} 1 & 0 & 0 \\
                          0 & -\xi/2 & 0 \\
                          0 & 0 & +\xi/2 
       \end{array}
\right] \;,
\label{Hdef2}
\end{equation}
where we have absorbed the extra $b$-terms in the $(1,1)$ element into $a$.

The extra $\xi$-dependent contribution in Eq.~(\ref{Hdef2}) 
can manifest itself when $a>|\delta m^2_{31}|$
(\textit{i.e.} $E\agt 10\,\mathrm{GeV}$ for typical matter densities in the Earth)
in the $\nu_\mu$ and $\bar{\nu}_\mu$ survival probabilities as \cite{HOT}
\begin{eqnarray}
{P}(\nu_\mu\rightarrow\nu_\mu) 
& \approx & 1-\sin^2\left(2\theta_{23} - \frac{a\xi}{\delta m^2_{31}}\right)\sin^2\dfrac{{\Delta}}{2}\;, \cr
{P}(\bar{\nu}_\mu \rightarrow \bar{\nu}_\mu)
& \approx & 1-\sin^2\left(2\theta_{23} + \frac{a\xi}{\delta m^2_{31}}\right)\sin^2\dfrac{{\Delta}}{2}\;,
\end{eqnarray}
where 
\begin{equation}
\Delta \approx \Delta_{31} c_{13}^2 - \Delta_{21} c_{12}^2\;,\qquad
\Delta_{ij} = \dfrac{\delta m^2_{ij}}{2E}L\;,\qquad
c_{ij} = \cos\theta_{ij}\;,
\end{equation}
and the CP violating phase $\delta$ has been set to zero.
As is evident from these expressions, the small shift due to $\xi$ will be invisible if
the value of $\sin^2 2\theta_{23}$ is too close to one.
However, if the value of $\sin^2 2\theta_{23}$ is as low as 
$\sin^2 2\theta_{23}=0.92$ (the current 90\% lower bound \cite{Ashie:2005ik}), and if
$\xi$ is as large as $\xi =0.025$ (the central value from CHARM/CHARM~II \cite{CHARM}),
then the shift in the survival probability at the first oscillation dip can be as large as $\sim 40\%$.
If the Fermilab-NUMI beam in its high-energy mode \cite{NUMI} 
were aimed at a declination angle of $46^\circ$ toward the planned 
Hyper-Kamiokande detector \cite{HyperK} in Kamioka, Japan
(baseline 9120~km), such a shift would be visible after just one year of data taking, assuming
a Mega-ton fiducial volume and 100\% efficiency.
The absence of any shift after 5 years of data taking would constrain $\xi$ to \cite{HOT}
\begin{equation}
|\xi| \le \xi_0 \equiv 0.005\;,
\label{xi_bound}
\end{equation}
at the 99\% confidence level.

In this paper, we look at how this potential limit on $\xi$ would translate into constraints on 
new physics, in particular, on the couplings and masses of new particles.
As mentioned above, the models must be those that distinguish among different generations.
We consider the following four classes of models: 
\begin{enumerate}
\item Models with a generation distinguishing $Z'$ boson. This class includes gauged $L_e-L_\mu$, gauged $L_e-L_\tau$, gauged $B-\alpha L_e-\beta L_\mu -\gamma L_\tau$, and topcolor assisted technicolor.
\item Models with leptoquarks (scalar and vector). This class includes various Grand Unification Theory (GUT) models and extended technicolor (ETC).
\item The Supersymmetric Standard Model with R-parity violation.
\item Extended Higgs models. This class includes the Babu model, the Zee model, and various models with triplet Higgs, as well as the generation distinguishing $Z'$ models listed above.
\end{enumerate}
These classes will be discussed one by one in sections II through V.
The constraints on these models will be compared with existing ones from LEP/SLD, the Tevatron, and other low energy experiments,
and with those expected from direct searches for the new particles at the LHC.
Concluding remarks will be presented in section VI.


\section{Models with an extra $Z'$ boson}
\label{ZprimeSection}


$Z^{\prime}$ generically refers to any electrically neutral gauge boson corresponding to a 
flavor-diagonal generator of some new gauge group.   
Here, we are interested in models in which the $Z'$ couples differently to
different generations.
The models we will consider are 
(A) gauged $L_e-L_\mu$ and $L_e-L_\tau$,
(B) gauged $B-\alpha L_e - \beta L_\mu - \gamma L_\tau$, with
$\alpha+\beta+\gamma = 3$, and
(C) topcolor assisted technicolor.

\subsection{Gauged $L_e-L_\mu$ and $L_e-L_\tau$}

\begin{figure}[ht]
\centering
    \begin{picture}(400,120)(-100,-80) 
    \SetWidth{1}
    \SetScale{1}  
    \SetColor{Black}
    
    \ArrowLine(-80,-40)(0,-40)
    \ArrowLine(0,-40)(80,-40)
    \Photon(0,-40)(0,40){7.5}{6}
    \ArrowLine(-80,40)(0,40)
    \ArrowLine(0,40)(80,40)
    \Vertex(0,40){2}
    \Vertex(0,-40){2}       
    \Text(-70,50)[]{$\nu_\mu$}
    \Text(70,50)[]{$\nu_\mu$}
    \Text(-70,-50)[]{$e$}
    \Text(70,-50)[]{$e$}
    \Text(17,0)[]{$Z^{\prime}$}
    \Text(0,50)[]{$-ig_{Z'}$}
    \Text(0,-50)[]{$+ig_{Z'}$}
    \Text(0,-75)[]{$(a)$}

    \SetOffset(200,0)
    \ArrowLine(-80,-40)(0,-40)
    \ArrowLine(0,-40)(80,-40)
    \Photon(0,-40)(0,40){7.5}{6}
    \ArrowLine(-80,40)(0,40)
    \ArrowLine(0,40)(80,40)
    \Vertex(0,40){2}
    \Vertex(0,-40){2}       
    \Text(-70,50)[]{$\nu_\tau$}
    \Text(70,50)[]{$\nu_\tau$}
    \Text(-70,-50)[]{$e$}
    \Text(70,-50)[]{$e$}
    \Text(17,0)[]{$Z^{\prime}$}
    \Text(0,50)[]{$-ig_{Z'}$}
    \Text(0,-50)[]{$+ig_{Z'}$}
    \Text(0,-75)[]{$(b)$}
    
    \end{picture}
	\caption{Diagrams that contribute to neutrino oscillation matter effects in 
	(a) the gauged $L_e-L_\mu$ model, and 
	(b) the gauged $L_e-L_\tau$ model.}
	\label{L1minusL2figure}
\end{figure}
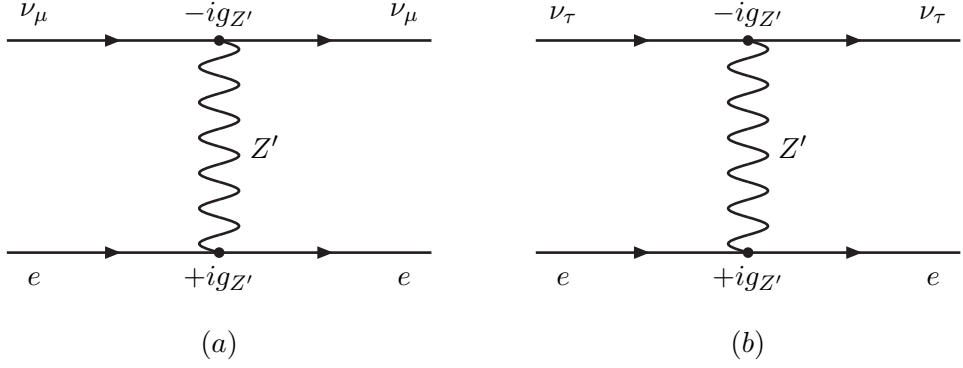

In Ref.~\cite{L1minusL2}, it was pointed out that the charges $L_e-L_\mu$, $L_e-L_\tau$, and $L_\mu-L_\tau$ are anomaly free within the particle content of the
Standard Model, and therefore can be gauged.  Models with these symmetries are
recently receiving renewed attention in attempts to explain the large mixing angles observed in the neutrino sector \cite{LmuMinusLtau}.
Of these, gauged $L_e-L_\mu$ and $L_e-L_\tau$ affect neutrino oscillation 
in matter.
These models necessarily possess a Higgs sector which also distinguishes among
different lepton generations \cite{Foot:2005uc}, but we will only consider the
effect of the the extra gauge boson in this section and relegate the effect of the Higgs sector to a more generic discussion in section~V.

The interaction Lagrangian for gauged $L_e-L_\ell$ ($\ell=\mu$ or $\tau$) is 
given by
\begin{equation}
\mathcal{L} = g_{Z'}
\left(\, \overline{e}\gamma^\mu e - \overline{\ell}\gamma^\mu \ell 
+ \overline{\nu_{eL}}\gamma^\mu \nu_{eL} 
- \overline{\nu_{\ell L}}\gamma^\mu \nu_{\ell L}
\,\right)
Z'_\mu \;.
\end{equation}
The diagrams that affect neutrino propagation in matter are shown
in Fig.~\ref{L1minusL2figure}.  (The exchange of the $Z'$ between the $\nu_e$ 
and the electrons do not lead to new matter effects.) 
The forward scattering amplitude of the left-handed neutrino
$\nu_{\ell L}$ ($\ell=\mu,\tau$) is
\begin{equation}
i\mathcal{M} = (ig_{Z'})(-ig_{Z'})
\bra{\nu_{\ell L}}\overline{\nu_{\ell L}}\gamma^\mu\nu_{\ell L}
\ket{\nu_{\ell L}}
\left(\dfrac{ig_{\mu\nu}}{M_{Z'}^2}\right)
\bra{e}\overline{e}\gamma^\nu e\ket{e} \;.
\end{equation}
The electrons in matter are non-relativistic, 
so only the time-like components of the currents need to be considered.
Replacing $\bra{e}\overline{e}\gamma^0 e\ket{e} = \bra{e}e^\dagger e\ket{e}$ with
$N_e$, the number density of electrons in matter, and
$\bra{\nu_{\ell L}}\overline{\nu_{\ell L}}\gamma^0\nu_{\ell L}\ket{\nu_{\ell L}}
= \bra{\nu_{\ell L}}\nu_{\ell L}^\dagger \nu_{\ell L}\ket{\nu_{\ell L}}$ with
$\phi_{\nu_\ell}^\dagger\phi_{\nu_\ell}$, where $\phi_{\nu_\ell}$ is the
wave function of the left-handed neutrino $\nu_{\ell L}$, we obtain
\begin{equation}
i\mathcal{M} 
\;=\; i\dfrac{g_{Z'}^2}{M_{Z'}^2}
\left(\phi_{\nu_{\ell}}^\dagger\phi_{\nu_\ell}\right) N_e
\;\equiv\; -iV_{\nu_\ell}\left(\phi_{\nu_{\ell}}^\dagger\phi_{\nu_\ell}\right)\;.
\end{equation}
Therefore, the effective potential felt by the neutrinos as they traverse matter
can be identified as
\begin{equation}
V_{\nu_\ell} = -\dfrac{g_{Z'}^2}{M_{Z'}^2}\,N_e\;.
\end{equation}
The effective $\xi$'s for the $L_e-L_\mu$ and $L_e-L_\tau$ cases are
\begin{eqnarray}
\xi_{L_e-L_\mu} 
& = & -\dfrac{V_{\nu_\mu}}{V_{CC}}
\;=\; +4\,\dfrac{(g_{Z'}^2/M_{Z'}^2)}{(g^2/M_W^2)}
\;=\; +\dfrac{1}{\sqrt{2}G_F}\left(\dfrac{g_{Z'}}{M_{Z'}}\right)^2\;,\cr
\xi_{L_e-L_\tau} 
& = & +\dfrac{V_{\nu_\tau}}{V_{CC}}
\;=\; -4\,\dfrac{(g_{Z'}^2/M_{Z'}^2)}{(g^2/M_W^2)}
\;=\; -\dfrac{1}{\sqrt{2}G_F}\left(\dfrac{g_{Z'}}{M_{Z'}}\right)^2\;.
\end{eqnarray}
Ignoring potential contributions from the Higgs sector, 
a bound on $\xi$ of $|\xi|\le \xi_0 = 0.005$ from Eq.~(\ref{xi_bound}) translates into:
\begin{equation}
\dfrac{M_{Z'}}{g_{Z'}} \ge \sqrt{\dfrac{1}{\sqrt{2}G_F \xi_0}} \approx 3500\,\mathrm{GeV}\;,
\label{L1minusL2bound}
\end{equation}
for both the $L_e-L_\mu$ and $L_e-L_\tau$ cases.

\begin{table}[t]
\begin{tabular}{|r||c|c|c||c|}
\hline
& \ $\Lambda_-$ (TeV) from\ \ 
& \ $\Lambda_+$ (TeV) from\ \ 
& \ $\Lambda_+$ (TeV) from\ \ 
& \\
& \ $e^+e^-\rightarrow e^+e^-$\ \ 
& \ $e^+e^-\rightarrow \mu^+\mu^-$\ \
& \ $e^+e^-\rightarrow \tau^+\tau^-$\ \
& \ Reference\ \ \\
\hline
\ L3\ \ & $10.1$ & $14.4$ & $\phantom{0}7.6$ & \cite{Acciarri:2000uh} \\
\ OPAL\ \ & $10.6$ & $12.7$ & $\phantom{0}8.6$ & \cite{Abbiendi:2003dh} \\
\ DELPHI\ \ & $13.9$ & $12.2$ & $15.8$ & \cite{Abdallah:2005ph} \\
\ ALEPH\ \ & $12.5$ & $10.5$ & $12.8$ & \cite{Schael:2006wu} \\
\hline
\end{tabular}
\caption{The 95\% confidence level lower bounds on the compositeness scale 
$\Lambda^\pm$ (TeV) from leptonic LEP/LEP2 data.
Dividing by $\sqrt{4\pi}$ converts these limits to those on $(M_{Z'}/g_{z'})$.}
\label{LEP2data}
\end{table}

The $Z'$ in gauged $L_e-L_\ell$ ($\ell=\mu,\tau$) cannot be sought for at the LHC since they
only couple to leptons.  However, they can be produced in $e^+e^-$ collisions
and subsequently decay into $e^+e^-$ or $\ell^+\ell^-$ pairs, and stringent contraints
already exist from LEP/LEP2.  The exchange of the $Z'$ induces the following effective 
four-fermion interactions, relevant to $e^+e^-$ colliders,
among the charged leptons at energies far below the $Z'$ mass:
\begin{equation}
\mathcal{L} = 
-\dfrac{g_{Z'}^2}{2M_{Z'}^2}
\left(\overline{e}\gamma_\mu e\right)\left(\overline{e}\gamma^\mu e\right)
+\dfrac{g_{z'}^2}{M_{Z'}^2}
\left(\overline{e}\gamma_\mu e\right)\left(\overline{\ell}\gamma^\mu \ell\right)
\;.
\end{equation}
The LEP collaborations fit their data to
\begin{equation}
\mathcal{L} = 
-\dfrac{4\pi}{2\Lambda_-^2}
\left(\overline{e}\gamma_\mu e\right)\left(\overline{e}\gamma^\mu e\right)
+\dfrac{4\pi}{\Lambda_+^2}
\left(\overline{e}\gamma_\mu e\right)\left(\overline{\ell}\gamma^\mu \ell\right) \;,
\end{equation}
with the 95\% confidence limits on $\Lambda_\pm$ shown in Table~\ref{LEP2data}.
The strongest constraint for the $L_e-L_\mu$ case comes from the $e^+e^-\rightarrow\mu^+\mu^-$
channel of L3, which translates to
\begin{equation}
\dfrac{M_{Z'}}{g_{Z'}} \ge 4.1\,\mathrm{TeV}\;, 
\end{equation}
while that for the $L_e-L_\tau$ case comes from the $e^+e^-\rightarrow\tau^+\tau^-$ channel
of DELPHI, which translates to
\begin{equation}
\dfrac{M_{Z'}}{g_{Z'}} \ge 4.5\,\mathrm{TeV}\;.
\end{equation}
Though these are the 95\% confidence limits while that given in Eq.~(\ref{L1minusL2bound})
is the 99\% limit, it is clear that the bound on $\xi$ will not lead to any improvement of
already existing bounds from LEP/LEP2.

\subsection{Gauged $B-(\alpha L_e + \beta L_\mu + \gamma L_\tau)$}

\begin{figure}[t]
\centering
    \begin{picture}(400,120)(-100,-80) 
    \SetWidth{1}
    \SetScale{1}  
    \SetColor{Black}
    
    \ArrowLine(-80,-40)(0,-40)
    \ArrowLine(0,-40)(80,-40)
    \Photon(0,-40)(0,40){7.5}{6}
    \ArrowLine(-80,40)(0,40)
    \ArrowLine(0,40)(80,40)
    \Vertex(0,40){2}
    \Vertex(0,-40){2}       
    \Text(-70,50)[]{$\nu_\ell$}
    \Text(70,50)[]{$\nu_\ell$}
    \Text(-70,-50)[]{$f$}
    \Text(70,-50)[]{$f$}
    \Text(17,0)[]{$Z^{\prime}$}
    \Text(0,50)[]{$+ig_{Z'} X_{\nu_\ell}$}
    \Text(0,-50)[]{$+ig_{Z'} X_f$}
    \Text(0,-75)[]{$(a)$}

    \SetOffset(200,0)
    \ArrowLine(-80,-40)(0,-40)
    \ArrowLine(0,-40)(80,-40)
    \Photon(0,-40)(0,40){7.5}{6}
    \ArrowLine(-80,40)(0,40)
    \ArrowLine(0,40)(80,40)
    \Vertex(0,40){2}
    \Vertex(0,-40){2}       
    \Text(-70,50)[]{$\nu_\tau$}
    \Text(70,50)[]{$\nu_\tau$}
    \Text(-70,-50)[]{$f$}
    \Text(70,-50)[]{$f$}
    \Text(17,0)[]{$Z^{\prime}$}
    \Text(0,50)[]{$+\frac{i}{2}\,g'\cot\theta_1$}
    \Text(0,-50)[]{$-ig'Y_f\tan\theta_1$}
    \Text(0,-75)[]{$(b)$}
    
    \end{picture}
	\caption{Diagrams that contribute to neutrino oscillation matter effects in 
	(a) the gauged $X=B-\alpha L_e-\beta L_\mu - \gamma L_\tau$ model, 
	$\ell=\{e,\mu,\tau\}$, $f=\{u,d,e\}$, and 
	(b) topcolor assisted technicolor, $f = \{ u_L, u_R, d_L, d_R, e_L, e_R\}$.}
	\label{Z`}
\end{figure}
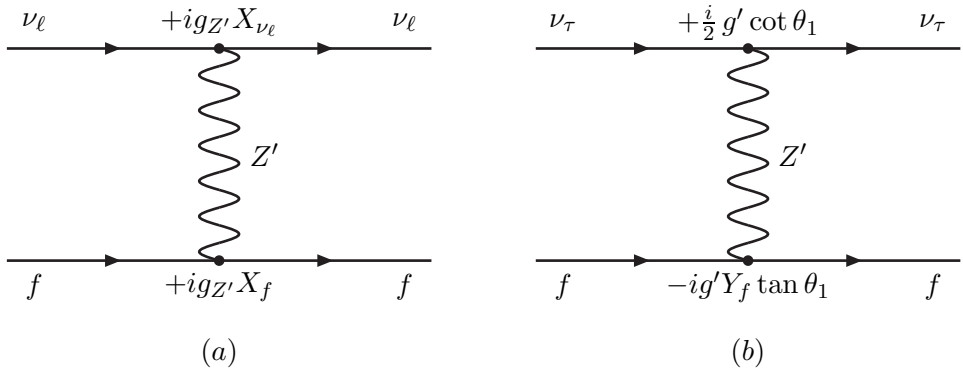

In Refs.~\cite{Bminus3Ltau,Bminus3Le,Bminus3over2LtauplusLmu,Chang:2000xy}, 
extensions of the SM gauge group to 
$SU(3)_C\times SU(2)_L\times U(1)_Y\times U(1)_X$
with $X=B-(\alpha L_e+\beta L_\mu+\gamma L_\tau)$ were considered.
Again, the motivation was to explain the observed pattern of neutrino masses and mixings. 
The cases 
$(\alpha,\beta,\gamma)=(0,0,3)$, $(3,0,0)$, and $(0,\frac{3}{2},\frac{3}{2})$
were considered, respectively, in Refs.~\cite{Bminus3Ltau}, \cite{Bminus3Le}, 
and \cite{Bminus3over2LtauplusLmu}.
In all cases, the condition 
\begin{equation} \label{anomaly_cancellation}
\alpha + \beta + \gamma = 3
\end{equation}
is required for anomaly
cancellation within the SM plus right-handed neutrinos\footnote{Only the right-handed neutrinos with non-zero $X$ charge need to be included for anomaly cancellation.}.
When $\alpha\neq\beta\neq\gamma$, the $U(1)_X$ gauge boson, \textit{i.e.} the $Z'$,
couples to the three lepton generations differently, 
and can lead to extra neutrino oscillation matter effects.
As in the gauged $L_e-L_\ell$ case,
the Higgs sectors of these models also necessarily 
distinguish among the lepton generations,
but we relegate the discussion of their effects to section~V.

For generic values of $(\alpha,\beta,\gamma)$, 
the $Z'$ couples to the quarks and leptons as
\begin{equation}
\mathcal{L}_{Z'} = g_{Z'}^{\phantom{\mu}} J_{X}^\mu Z_\mu'\;,
\label{LZ`}
\end{equation}
where 
\begin{eqnarray}
J_{X}^\mu
& = & \sum_{f} X_f (\bar{f}\gamma^\mu f) \cr
& = & 
\dfrac{1}{3} \sum_{q} \left(\,\bar{q}\gamma^\mu q\,\right) 
-\alpha \left(\, \bar{e}\gamma^\mu e + \overline{{\nu}_{e}}\gamma^\mu\nu_{e} \,\right)
-\beta  \left(\, \bar{\mu}\gamma^\mu \mu + \overline{{\nu}_{\mu}} \gamma^\mu \nu_{\mu} \,\right)
-\gamma \left(\, \bar{\tau}\gamma^\mu \tau + \overline{{\nu}_{\tau}} \gamma^\mu \nu_{\tau} \,\right)
\;.\cr
& &
\end{eqnarray}
The forward scattering amplitude of the left-handed 
neutrino $\nu_{\ell L}$ ($\ell=e,\mu,\tau$) on matter fermion $F$ ($F=p,n,e$) 
due to $Z'$-exchange (\textit{cf.} Fig.~\ref{Z`}a) is
\begin{eqnarray}
i\mathcal{M}_F\;=\;(+ig_{Z'}X_{\nu_\ell})(+ig_{Z'})\,
\bra{\nu_{\ell L}}\overline{\nu_{\ell}}\gamma^{\mu}\nu_{\ell}\ket{\nu_{\ell L}}
\left( \dfrac{ig_{\mu\nu}}{M^2_{Z^{\prime}}} \right)
\bra{F}J_X^\nu\ket{F}\;.
\end{eqnarray}
Again, we can assume that the matter fermions are non-relativistic, so that only the
time-like components of the currents need be considered.
Then, we can make the replacements
\begin{eqnarray}
\bra{e}J_X^0\ket{e} 
& = & -\alpha\bra{e}e^\dagger e\ket{e} 
\;\rightarrow\; -\alpha N_e\;,\cr
\bra{p}J_X^0\ket{p} 
& = & \dfrac{1}{3}\bra{p}\left(u^\dagger u + d^\dagger d\right)\ket{p}
\;\rightarrow\; \dfrac{1}{3}\left(2N_p + N_p\right) \;=\; N_p\;,\cr
\bra{n}J_X^0\ket{n} 
& = & \dfrac{1}{3}\bra{n}\left(u^\dagger u + d^\dagger d\right)\ket{n}
\;\rightarrow\; \dfrac{1}{3}\left(N_n + 2N_n\right) \;=\; N_n\;,
\end{eqnarray}
and
\begin{equation}
\bra{\nu_{\ell L}}\overline{\nu_\ell}\gamma^0\nu_\ell\ket{\nu_{\ell L}}
\;=\; \bra{\nu_{\ell L}}\left(\nu_{\ell L}^\dagger\nu_{\ell L}^{\phantom{\dagger}} + \nu_{\ell R}^\dagger\nu_{\ell R}^{\phantom{\dagger}}\right)\ket{\nu_{\ell L}}
\;=\; \bra{\nu_{\ell L}}\nu_{\ell L}^\dagger\nu_{\ell L}^{\phantom{\dagger}}\ket{\nu_{\ell L}}
\;\rightarrow\; \phi_{\nu_\ell}^\dagger\phi_{\nu_\ell}^{\phantom{\dagger}}\;,
\end{equation}
which gives us
\begin{equation}
i\mathcal{M}_F\;=\;-i X_{\nu_\ell}\frac{g^2_{Z^{\prime}}}{M^2_{Z^{\prime}}} 
\left( \phi_{\nu_\ell}^{\dagger}\phi_{\nu_\ell}^{\phantom{\dagger}} \right)
(X_F N_F)  \;,
\end{equation} 
where we have defined $X_p=X_n=1$.
Summing over $F=p,n,e$, we find:
\begin{eqnarray}
i\mathcal{M} 
& = & i\sum_{F=p,n,e} \mathcal{M}_F  \cr
& = & -iX_{\nu_\ell} 
\dfrac{g_{Z'}^2}{M_{Z'}^2}
\left( \phi_{\nu_\ell}^{\dagger}\phi_{\nu_\ell}^{\phantom{\dagger}} \right)
\left(\; N_p + N_n -\alpha N_e \;\right)
\;=\; -i\,V_{\nu_\ell}
\left( \phi_{\nu_\ell}^{\dagger}\phi_{\nu_\ell}^{\phantom{\dagger}} \right) \;,
\cr & &
\end{eqnarray}
where
\begin{equation}
V_{\nu_\ell}\equiv +X_{\nu_\ell}
\frac{g^2_{Z^{\prime}}}{M^2_{Z^{\prime}}}
\left(N_n+N_p-\alpha N_e\right)
\end{equation}
can be identified as the effective potential experienced by the left-handed
neutrino $\nu_{\ell L}$ as it travels through matter. 
Since the Earth is electrically neutral and is mostly composed of lighter elements,
we can make the approximation $N_n\approx N_p = N_e \equiv N$, in which case
\begin{equation}
V_{\nu_\ell}\approx -X_{\nu_\ell}
\frac{g^2_{Z^{\prime}}}{M^2_{Z^{\prime}}}
(\alpha-2) N\;.
\end{equation}
The effective $\xi$ is then
\begin{equation}\label{xi}
\xi_{(\alpha,\beta,\gamma)} 
\;=\; \dfrac{V_{\nu_\tau}-V_{\nu_\mu}}{V_{CC}}
\;=\; -4(\alpha-2)(\beta-\gamma)\dfrac{(g_{Z'}/M_{Z'})^2}{(g/M_W)^2} \;.
\end{equation}
When $\alpha=2$, the contribution of the matter electrons is cancelled by those of the
matter nucleons and $\xi_{(2,\beta,\gamma)}$ vanishes, 
regardless of the values of $\beta$ and $\gamma$.
When $\beta=\gamma$,  the matter effects on $\nu_\mu$ and $\nu_\tau$
will be the same, again resulting in $\xi_{(\alpha,\beta,\beta)}=0$, regardless of the value of $\alpha$.

\begin{figure}[ht]
	\centering
		\includegraphics[width=8.1cm]{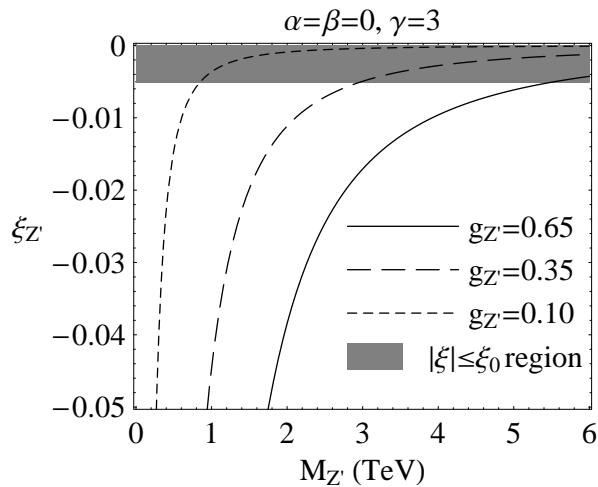}		
		\caption{$\xi_{Z^{\prime}}$ dependence on the $Z^{\prime}$ mass for the special case $\alpha = \beta=0$, $\gamma = 3$.}
	\label{Z`_result_xi_mass}
\end{figure}

In Fig.~\ref{Z`_result_xi_mass}, we plot the dependence of $\xi_{Z^{\prime}}$ on the 
$Z^{\prime}$ mass for selected values of $g_{Z'}$ for the case $\alpha = \beta=0$, $\gamma = 3$, 
namely, the $Z^\prime$ couples to $B-3L_\tau$.
In this case
\begin{equation}
\xi_{(0,0,3)}
\;=\; -24\,\dfrac{(g_{Z'}/M_{Z'})^2}{(g/M_W)^2}
\;=\; -\frac{6}{\sqrt{2}G_F}\, \left(\frac{g_{Z^{\prime}}}{M_{Z^{\prime}}}\right)^2\;.
\end{equation}
Ignoring the possible contribution of the Higgs sector,
a bound on $\xi$ of $|\xi|\le \xi_0 = 0.005$ from Eq.~(\ref{xi_bound}) translates into:
\begin{equation}
\dfrac{M_{Z'}}{g_{Z'}} \ge \sqrt{\dfrac{6}{\sqrt{2}G_F \xi_0}} \approx 8500\,\mathrm{GeV}\;.
\label{003ModelBound}
\end{equation}
%
\begin{figure}[ht]
	\centering
		\includegraphics[width=8.1cm]{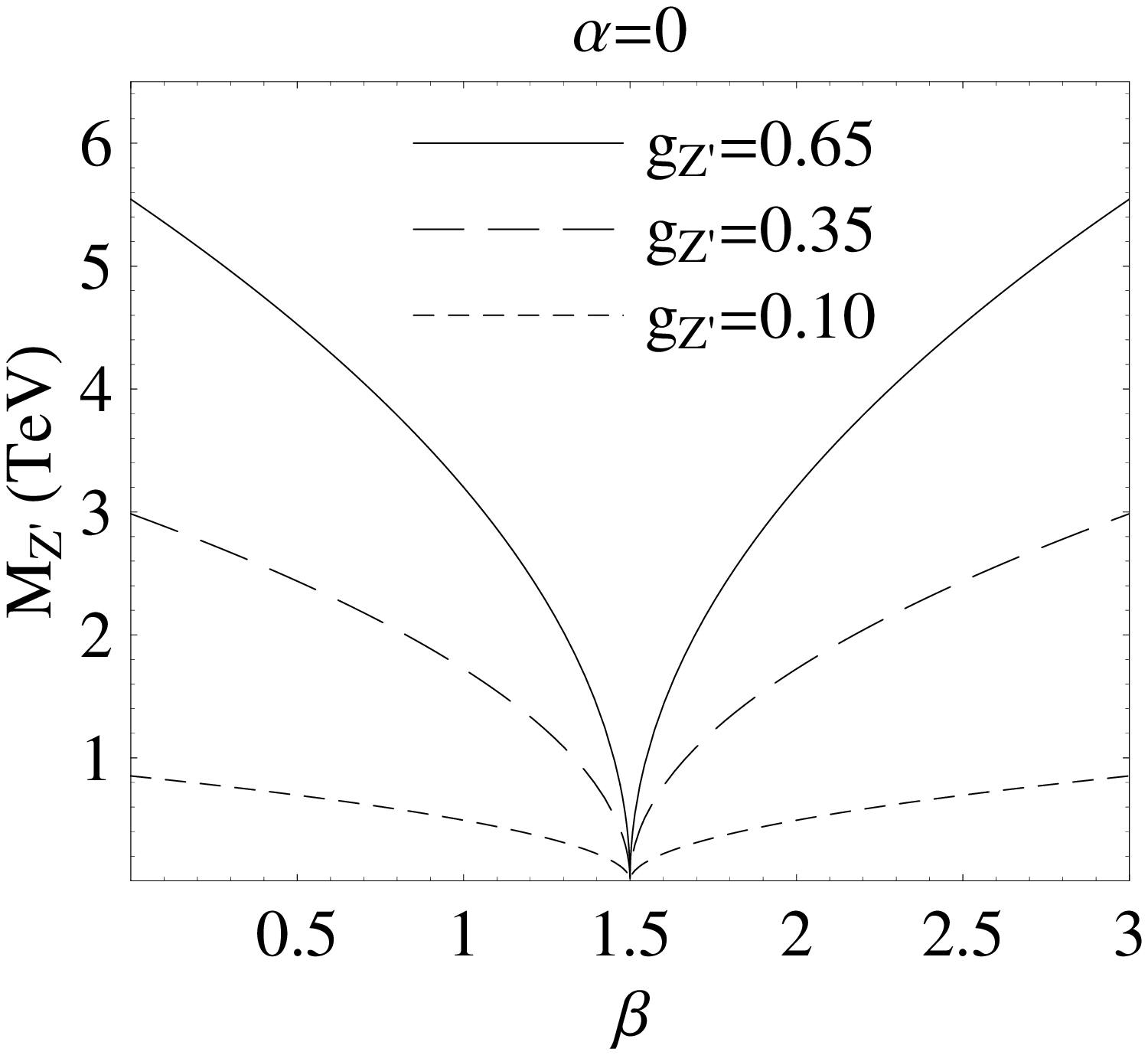}		
		\includegraphics[width=8.1cm]{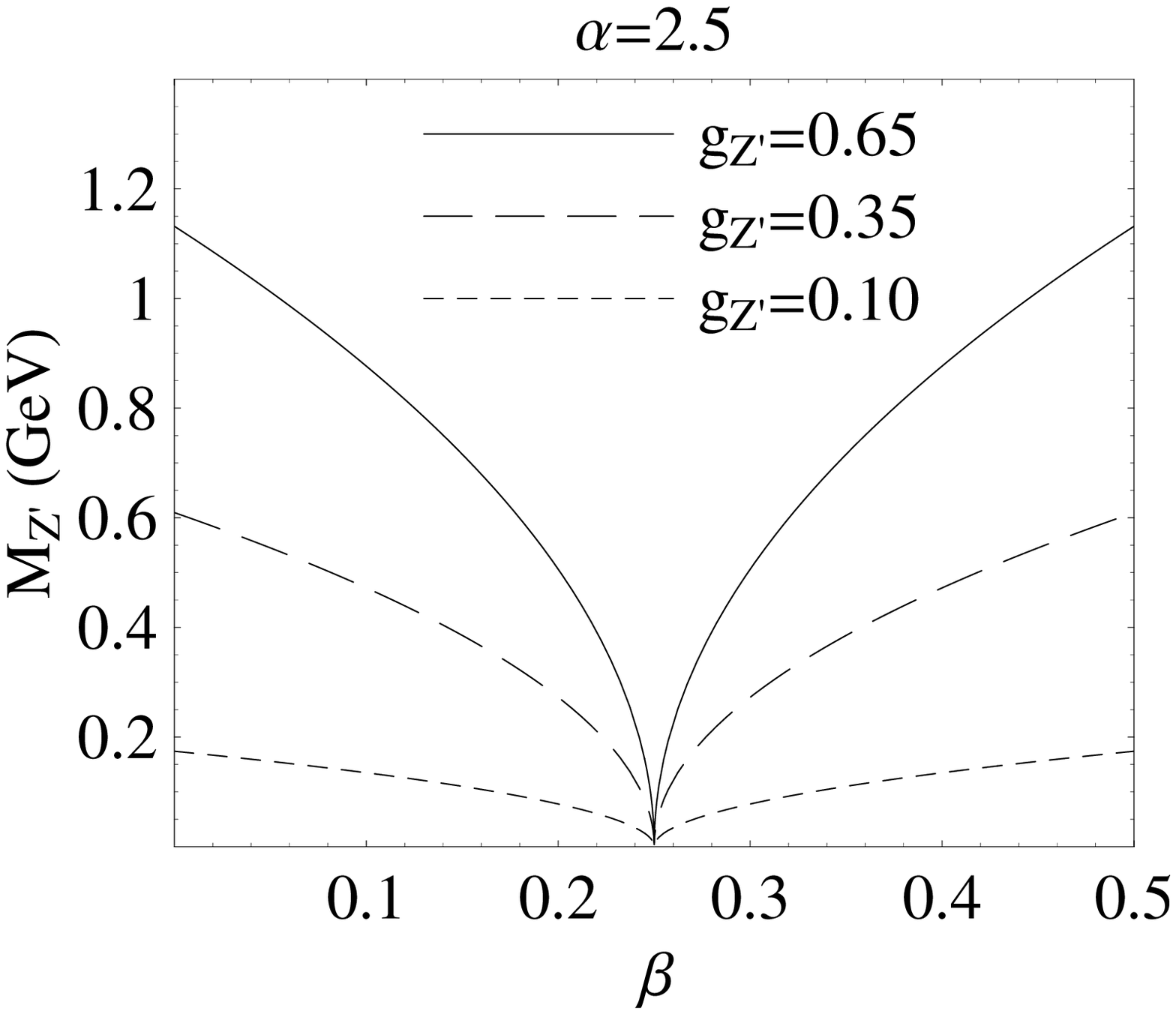}
	\caption{Lower bounds on $Z^{\prime}$ mass.}
	\label{Z`_result}
\end{figure}

\noindent
More generically, the bound on the $Z'$ mass is
\begin{equation} 
\dfrac{M_{Z^{\prime}}}{g_{Z^{\prime}}} \;\ge\; 
\sqrt{\frac{|(\alpha-2)(\beta-\gamma)|}{\sqrt{2}\,G_F \xi_0}}
\;\approx\; \sqrt{ |(\alpha-2)(\beta-\gamma)|}\times (3500\,\mathrm{GeV}) \;\;.
\end{equation}
This bound is plotted in Fig.~\ref{Z`_result} as a function of $\beta$ 
for three different values of $g_{Z'}$, and two different values of $\alpha$.
The value of $\gamma$ is fixed by the anomaly cancellation condition, Eq.~(\ref{anomaly_cancellation}),
to $\gamma = 3-\alpha-\beta$. 
The region of the $(\beta, M_{Z^{\prime}})$ parameter space below each curve will be excluded. 

\begin{center}
\begin{table}[tbp]
\begin{tabular}{|c||c|c|c|c|} 
\hline
& 
& \ 2$\sigma$ (95\%) limit from\ \
& \ 95\% limit from\ \
& \ limit from\ \
\\
$\;\;(\alpha,\beta,\gamma)\;\;$         
& $\;\; g_{Z^{\prime}} \;\;$
& \ LEP/SLD \cite{Chang:2000xy}\ \                  
& \ CDF \cite{Acosta:2005ij}/D0 \cite{Xuan:2005gw}\ \ 
& \ $|\xi|\le\xi_0$ (99\%)\ \     
\\
\hline\hline
$(0,0,3)$                    
& $0.65$ 
& $580\text{ GeV}$        
& $\sim 1\text{ TeV}$                    
& $5500\text{ GeV}$ 
\\
\cline{2-5}
& $0.35$                     
& $220\text{ GeV}$        
& $\sim 0.6\text{ TeV}$
& $3000\text{ GeV}$ 
\\
\hline
$\left(0,\frac{3}{2},\frac{3}{2}\right)$                          
& $0.65$                     
& $500\text{ GeV}$       
& $\phantom{\sim}880\text{ GeV}$
& ---                      
\\
\cline{2-5}
& $0.35$                     
& ---         
& $\phantom{\sim}470\text{ GeV}$
& ---                       
\\
\hline
\end{tabular}
\caption{Current and possible lower bounds on the $Z^{\prime}$ mass in gauged $B-\alpha L_3-\beta L_\mu-\gamma L_\tau$ models.}
\label{bounds_on_MZprime}
\end{table}
\end{center}

Let us now look at existing bounds.  
We limit our attention to the $\alpha=0$ case, i.e. the
$Z'$ couples to $B-\beta L_\mu - \gamma L_\tau$, with $\beta+\gamma=3$. 
In this case, the $Z'$ can be produced in $p\bar{p}$ collisions
and subsequently decay into $\mu^+\mu^-$ or $\tau^+\tau^-$ pairs.
The exchange of the $Z'$ in this case leads to the following
four-fermion interactions, relevant to $p\bar{p}$ colliders, 
between the charged leptons and the light quarks at energies way below the $Z'$ mass:
\begin{equation}
\mathcal{L} = 
+\dfrac{\beta g_{Z'}^2}{3M_{Z'}^2}
\left(\bar{u}\gamma^\mu u + \bar{d}\gamma^\mu d \right)
\left(\bar{\mu}\gamma_\mu\mu\right)
+\dfrac{\gamma g_{Z'}^2}{3M_{Z'}^2}
\left(\bar{u}\gamma^\mu u + \bar{d}\gamma^\mu d \right)
\left(\bar{\tau}\gamma_\mu\tau\right) \;.
\end{equation}
D0 has searched for the contact interaction
\begin{equation}
\mathcal{L} =
+\dfrac{4\pi}{\Lambda_+^2}
\left(\bar{u}\gamma^\mu u + \bar{d}\gamma^\mu d \right)
\left(\bar{\mu}\gamma_\mu\mu\right)
\end{equation}
in its dimuon production data \cite{Xuan:2005gw} and has set a 95\% confidence level limit of
\begin{equation}
\Lambda_+ \ge 6.88\,\mathrm{TeV}\;.
\end{equation}
This translates into
\begin{equation}
\dfrac{M_{Z'}}{g_{Z'}} \ge \sqrt{|\beta|}\times(1.1\,\mathrm{TeV})\;.
\end{equation}
CDF has searched for the production of a $Z'$ followed by its decay into
$\tau^+\tau^-$ pairs \cite{Acosta:2005ij} and has set a 95\% confidence level lower bound of
\begin{equation}
M_{Z'} \ge 400\,\mathrm{GeV}
\end{equation}
for a sequential $Z'$ (i.e. a $Z'$ with the exact same couplings to the fermions as the 
SM $Z$).  Rescaling to account for the difference in couplings, we estimate
\begin{equation}
\dfrac{M_{Z'}}{g_{Z'}} \agt \sqrt{|\gamma|}\times (1\,\mathrm{TeV})\;.
\end{equation}
Limits on this model also exist from a global analysis of loop effects in LEP/SLD data \cite{Chang:2000xy}, but they are weaker than the direct search limits from the Tevatron.
In Table~\ref{bounds_on_MZprime}, we compare the bounds from LEP/SLD, CDF/D0, and the potential
bounds from a measurement of $\xi$ for two choices of $(\alpha,\beta,\gamma)$, and
two choices for the value of $g_{Z'}$.
For the $(\alpha,\beta,\gamma)=(0,0,3)$ case, we can expect a significant improvement over current bounds.

The sensitivity of the LHC to $Z'$s has been analyzed assuming $Z'$ decay into $e^+e^-$ or
$\mu^+\mu^-$ pairs, or 2 jets \cite{LHC-TDR}. 
For a sequential $Z'$, the LHC is sensitive to masses as heavy as 5~TeV with
100~fb${}^{-1}$ of integrated luminosity.
The $Z'$ of the $(\alpha,\beta,\gamma)=(0,0,3)$ model, however, decays mostly into $\tau^+\tau^-$, which will not provide as clean a signal as decays into the lighter charged lepton pairs.
Ref.~\cite{Ma:1998dp} estimates that if $g_{Z'}\sim g'\approx 0.35$, then the
LHC reach will be up to about 1~TeV with 100~fb${}^{-1}$.
If this estimate is correct, the potential bound on $M_{Z'}$
from neutrino oscillation may be better than that from the LHC.
A complete detector analysis may show that the actual reach of the LHC is somewhat higher, but even then
we can expect the neutrino oscillation bound to be competitive with the LHC bound for the $(0,0,3)$ model.


\subsection{Topcolor Assisted Technicolor}

Another example of a model with a $Z'$ which distinguishes among different generations
is topcolor assisted technicolor \cite{TopTechni,TopTechni_limits}.
Models of this class are hybrids of topcolor and technicolor: the topcolor
interactions generate the large top-mass (and a fraction of the $W$ and $Z$ masses), while the technicolor interactions generate (the majority of ) the $W$ and $Z$ masses.
The models include a $Z'$ in the topcolor sector, the interactions of which 
helps the top to condense, but prevents the bottom from doing so also.
To extract the interactions of this $Z'$ relevant to our discussion, we need to look at the model in some detail.

Though there are several different versions of topcolor assisted technicolor, 
we consider here the simplest in which
the quarks and leptons transform under the gauge group
\begin{equation}
SU(3)_s \times SU(3)_w \times 
 U(1)_s \times  U(1)_w \times SU(2)_L
\end{equation}
with coupling constants $g_{3s}$, $g_{3w}$, $g_{1s}$, $g_{1w}$,
and $g$.  It is assumed that $g_{3s} \gg g_{3w}$ and
$g_{1s} \gg g_{1w}$.
$SU(2)_L$ is the usual weak-isospin gauge group of the SM with coupling constant $g$.
The charge assignments of the three generation of ordinary fermions
under these gauge groups are given in Table~\ref{CHARGE-ASSIGNMENTS}.
Note that each generation must transform non-trivially under
only one of the $SU(3)$'s and one of the $U(1)$'s, and that those
charges are the same as that of the SM color, and hypercharge $Y$
(normalized to $Q_{em}=I_3+Y$).
This ensures anomaly cancellation.

\begin{table}[t]
\begin{center}
\begin{tabular}{|c||c|c|c|c|c|}
\hline
           & $SU(3)_s$ & $SU(3)_w$ & $U(1)_s$ & $U(1)_w$ & $SU(2)_L$ \\
\hline\hline
$(t,b)_L$  &  3  &  1  & $\dfrac{1}{6}$  &   0   &   2   \\
\hline
$(t,b)_R$  &  3  &  1  & $\left(\dfrac{2}{3},-\dfrac{1}{3}\right)$
                                        &   0   &   1   \\
\hline
$(\nu_\tau,\tau^-)_L$
           &  1  &  1  & $-\dfrac{1}{2}$     &   0   &   2   \\
\hline
$\tau^-_R$ &  1  &  1  & $-1$     &   0   &   1   \\
\hline
$(c,s)_L$, 
$(u,d)_L$  &  1  &  3  &  0  & $\dfrac{1}{6}$   &   2   \\
\hline
$(c,s)_R$, 
$(u,d)_R$  &  1  &  3  &  0  & $\left(\dfrac{2}{3},-\dfrac{1}{3}\right)$
                                                &   1   \\
\hline
$(\nu_\mu,\mu^-)_L$, $(\nu_e,e^-)_L$
           &  1  &  1  &  0  & $-\dfrac{1}{2}$  &   2   \\
\hline
$\mu^-_R$, 
$e^-_R$    &  1  &  1  &  0  & $-1$  &   1   \\
\hline
\end{tabular}
\end{center}
\caption{Charge assignments of the ordinary fermions.
The $U(1)$ charges are equal to the SM hypercharges normalized to $Q_{em}=I_3+Y$.
}
\label{CHARGE-ASSIGNMENTS}
\end{table}

At scale $\Lambda\sim 1$~TeV, technicolor, which is included in the model to generate the
$W$ and $Z$ masses, is assumed to become strong and generate a condensate 
(of something which is left unspecified)
which breaks the two $SU(3)$'s and the two $U(1)$'s to their diagonal
subgroups:
\begin{equation}
SU(3)_s \times SU(3)_w \rightarrow SU(3)_c\;,\qquad
U(1)_s \times U(1)_w \rightarrow U(1)_Y\;,
\end{equation}
which we identify
with the usual SM color and hypercharge groups.
The massless unbroken SU(3) gauge bosons (the gluons $G_\mu^a$) and 
the massive broken SU(3) gauge bosons 
(the so called \textit{colorons} $C_\mu^a$) 
are related to the original $SU(3)_s\times SU(3)_w$ gauge fields
$X_{s\mu}^a$ and $X_{w\mu}^a$ by
\begin{eqnarray}
C_\mu & = & X_{s\mu} \cos\theta_3 - X_{w\mu} \sin\theta_3 \cr
G_\mu & = & X_{s\mu} \sin\theta_3 + X_{w\mu} \cos\theta_3
\end{eqnarray}
where we have suppressed the color indices, and
\begin{equation}
\tan\theta_3 = \frac{g_{3w}}{g_{3s}}\;.     
\end{equation}
The currents to which the gluons and colorons couple to are:
\begin{equation}
g_{3s}J_{3s}^\mu X_{s\mu} + g_{3w}J_{3w}^\mu X_{w\mu}
= g_3\left( \cot\theta_3 J_{3s}^\mu - \tan\theta_3 J_{3w}^\mu
     \right) C_\mu
+ g_3\left( J_{3s}^\mu + J_{3w}^\mu \right) G_\mu\;,
\end{equation}
where
\begin{equation}  
\frac{1}{g_3^2} = \frac{1}{g_{3s}^2} + \frac{1}{g_{3w}^2}\;.  
\end{equation}
Since the quarks carry only one of the $SU(3)$ charges, we can identify
\begin{equation}   
J_3^\mu = J_{3s}^\mu + J_{3w}^\mu   
\end{equation}
as the QCD color current, and $g_3$ as the QCD coupling constant.

Similarly, the massless unbroken U(1) gauge boson $B_\mu$ and 
the massive broken U(1) gauge boson $Z'_\mu$
are related to the original $U(1)_s\times U(1)_w$ gauge fields
$Y_{s\mu}$ and $Y_{w\mu}$ by
\begin{eqnarray}
Z'_\mu & = & Y_{s\mu} \cos\theta_1 - Y_{w\mu} \sin\theta_1 \cr
B_\mu  & = & Y_{s\mu} \sin\theta_1 + Y_{w\mu} \cos\theta_1
\end{eqnarray}
where
\begin{equation} 
\tan\theta_1 = \frac{g_{1w}}{g_{1s}}\;.
\label{theta1}     
\end{equation}
The currents to which the $B_\mu$ and $Z'_\mu$ couple to are:
\begin{equation}
g_{1s}J_{1s}^\mu Y_{s\mu} + g_{1w}J_{1w}^\mu Y_{w\mu}
= g_1\left( \cot\theta_1 J_{1s}^\mu - \tan\theta_1 J_{1w}^\mu
     \right) Z'_\mu
+ g_1\left( J_{1s}^\mu + J_{1w}^\mu \right) B_\mu\;,
\end{equation}
where
\begin{equation} 
\frac{1}{g_1^2} = \frac{1}{g_{1s}^2} + \frac{1}{g_{1w}^2}\;.  
\end{equation}
Again,
since the fermions carry only one of the $U(1)$ charges, we can identify
\begin{equation}   
J_1^\mu = J_{1s}^\mu + J_{1w}^\mu   
\end{equation}
as the SM hypercharge current, and $g_1$ as the SM hypercharge
coupling constant $g'$.
Note that the interactions of the colorons and the $Z'$ with the third generation fermions are strong,
while their interactions with the first and second generation fermions are weak. This results in the
formation of a top-condensate which accounts for the large mass of the top quark.\footnote{
The $Z'$-exchange interaction in the $t\bar{t}$ channel is attractive, but that in the $b\bar{b}$
channel is repulsive.  This repulsion is assumed to be strong enough to counter the attraction
due to the colorons and prevent the bottom from condensing.}

Therefore, the interaction of the $Z'$ in this model with the quarks and leptons is given by
\begin{equation}
\mathcal{L} =
g'\left( \cot\theta_1 J_{1s}^\mu - \tan\theta_1 J_{1w}^\mu \right) Z'_{\mu}\;,
\end{equation}
where $g'$ is the SM hypercharge coupling, and
\begin{eqnarray}
J_{1s}^\mu & = &
\dfrac{1}{6}\left(\bar{t}_L\gamma^\mu t_L + \bar{b}_L\gamma^\mu b_L\right)
+\dfrac{2}{3}\bar{t}_R\gamma^\mu t_R -\dfrac{1}{3}\bar{b}_R\gamma^\mu b_R
-\dfrac{1}{2}\left(\bar{\tau}_L\gamma^\mu \tau_L + \bar{\nu}_{\tau L}\gamma^\mu \nu_{\tau L}\right)
-\bar{\tau}_R\gamma^\mu \tau_R \;,\cr
J_{1w}^\mu & = &
\dfrac{1}{6}\left(\bar{c}_L\gamma^\mu c_L + \bar{s}_L\gamma^\mu s_L\right)
+\dfrac{2}{3}\bar{c}_R\gamma^\mu c_R -\dfrac{1}{3}\bar{s}_R\gamma^\mu s_R
-\dfrac{1}{2}\left(\bar{\mu}_L\gamma^\mu \mu_L + \bar{\nu}_{\mu L}\gamma^\mu \nu_{\mu L}\right)
-\bar{\mu}_R\gamma^\mu \mu_R \cr
& + &
\dfrac{1}{6}\left(\bar{u}_L\gamma^\mu u_L + \bar{d}_L\gamma^\mu d_L\right)
+\dfrac{2}{3}\bar{u}_R\gamma^\mu u_R -\dfrac{1}{3}\bar{d}_R\gamma^\mu d_R
-\dfrac{1}{2}\left(\bar{e}_L\gamma^\mu e_L + \bar{\nu}_{e L}\gamma^\mu \nu_{e L}\right)
-\bar{e}_R\gamma^\mu e_R \;.\cr
& &
\end{eqnarray}
The exchange of the $Z'$ leads to the current-current interaction
\begin{equation}
\dfrac{1}{2}\left( \cot\theta_1 J_{1s} - \tan\theta_1 J_{1w} \right)
\left( \cot\theta_1 J_{1s} - \tan\theta_1 J_{1w} \right)
\;,
\label{Current_Current}
\end{equation} 
the $J_{1s}J_{1s}$ part of which does not contribute to neutrino oscillations on the Earth, 
while the $J_{1w}J_{1w}$ part is suppressed relative to the $J_{1w}J_{1s}$ part by
a factor of $\tan^2\theta_1 \ll 1$.
Therefore, we only need to consider the $J_{1s}J_{1w}$ interaction which only affects the propagation of $\nu_{\tau L}$ (\textit{cf.} Fig.~\ref{Z`}b). 
The forward scattering amplitude of $\nu_{\tau L}$ against fermion $F=p,n,e$ is given by
\begin{eqnarray} 
i\mathcal{M} 
& = & (-ig^{\prime}\cot\theta_1)(+ig^\prime\tan\theta_1)
\bra{\nu_{\tau L}} \left( -\dfrac{1}{2}\,\overline{\nu_{\tau}}\gamma^{\mu}P_L\nu_{\tau} \right) \ket{\nu_{\tau L}}
\;\dfrac{ig_{\mu\nu}}{M^2_{Z^{\prime}}} \cr
& & \times
\bra{F} \left[
\overline{u}\gamma^{\nu}\left(\dfrac{1}{6}P_L+\dfrac{2}{3}P_R\right)u 
+ \overline{d}\gamma^{\nu}\left(\dfrac{1}{6}P_L-\dfrac{1}{3}P_R\right)d+
\overline{e}\gamma^{\nu}\left(-\dfrac{1}{2}P_L-P_R\right)e 
\right] \ket{F} \cr
& \rightarrow & 
-\dfrac{ig^{\prime 2}}{2M_{Z^{\prime}}^2}
\left( \phi_{\nu_{\tau}}^{\dagger}\phi_{\nu_{\tau}}^{\phantom{\dagger}} \right)
\left[
\dfrac{1}{2} \left( \dfrac{1}{6}+\dfrac{2}{3} \right) ( 2N_p + N_n )+
\dfrac{1}{2} \left( \dfrac{1}{6}-\dfrac{1}{3} \right) (N_p+2N_n) +
\dfrac{1}{2} \left(-\dfrac{1}{2}-1\right)N_e
\right] \cr
& = & 
-\dfrac{ig^{\prime 2}}{2M_{Z^{\prime}}^2}
\left( \phi_{\nu_{\tau}}^{\dagger}\phi_{\nu_{\tau}}^{\phantom{\dagger}} \right)
\left( \dfrac{3}{4}N_p +\dfrac{1}{4}N_n - \dfrac{3}{4} N_e \right) \cr
& = & 
-\dfrac{ig^{\prime 2}}{8M_{Z^{\prime}}^2}
\left( \phi_{\nu_{\tau}}^{\dagger}\phi_{\nu_{\tau}}^{\phantom{\dagger}} \right)
N_n \cr
& \approx & 
- i\left(\frac{\displaystyle g^{\prime 2}}{\displaystyle M_{Z^{\prime}}^2 }\right)
\frac{\displaystyle N}{\displaystyle 8}
\left( \phi_{\nu_{\tau}}^{\dagger}\phi_{\nu_{\tau}}^{\phantom{\dagger}} \right)
\;=\;-iV_{\nu_{\tau}}\left( \phi_{\nu_{\tau}}^{\dagger}\phi_{\nu_{\tau}}^{\phantom{\dagger}} \right)\;.
\label{TAT matrix element}
\end{eqnarray}
Note that the angle $\theta_1$ has vanished from this expression and the only 
unknown parameter here is the $Z'$ mass.

\begin{figure}[t]
	\centering
		\includegraphics[width=8.1cm]{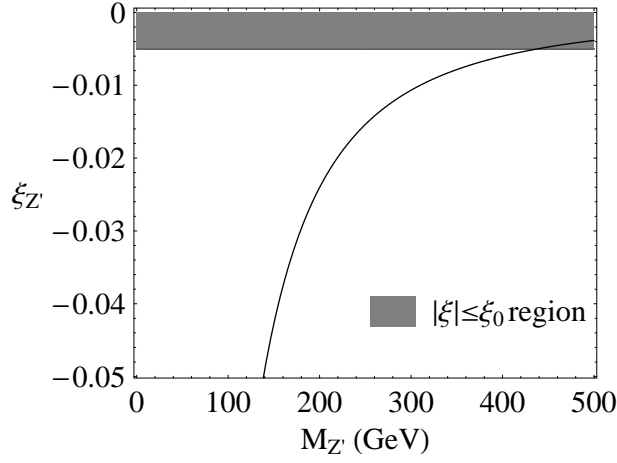}		
		\caption{$\xi_{TT}$ dependence on the $Z^{\prime}$ mass in the top color assisted technicolor model.}
	\label{Z`_result_TT}
\end{figure}

The effective potentials felt by the different neutrino flavors are
\begin{equation}
V_{\nu_e}    \;=\; V_{\nu_\mu}  \;=\; 0 \;, \qquad
V_{\nu_\tau} \;=\; +\dfrac{N}{8}\dfrac{g^{\prime 2}}{M_{Z'}^2} \;,
\end{equation}
and the effectivie $\xi$ is
\begin{equation}
\xi_{TT} 
= \frac{V_{\nu_\tau} - V_{\nu_\mu}}{V_{CC}}
= \frac{1}{2}\,\dfrac{(g'/M_{Z'})^2}{(g/M_W)^2}
= \frac{1}{2}\tan^2\theta_W\,\dfrac{M_W^2}{M_{Z'}^2}
= \frac{1}{2}\sin^2\theta_W\,\dfrac{M_Z^2}{M_{Z'}^2} 
\;.
\end{equation}
The dependence of
$\xi_{TT}$ on the  $Z^{\prime}$ mass is shown in Fig.~\ref{Z`_result_TT}. 
The limit $|\xi_{TT}|\le \xi_{0} = 0.005$ in this case translates to:
\begin{equation}
M_{Z^{\prime}} \ge 
M_Z \sqrt{ \dfrac{\sin^2\theta_W}{2\xi_0} }
\approx 440\,\mathrm{GeV}\;.
\end{equation}
This potential limit from the measurement of $\xi$ 
is much weaker than what is already available from precision electroweak data 
\cite{TopTechni_limits}, or from the direct search for 
$p\bar{p}\rightarrow Z'X \rightarrow \tau^+\tau^-X$ at CDF 
mentioned earlier \cite{Acosta:2005ij}.

\section{Generation Non-Diagonal Leptoquarks}

Leptoquarks are particles carrying both baryon number $B$, and lepton number $L$. 
They occur in various extensions of the SM such as Grand Unification Theories (GUT's) 
or Extended Technicolor (ETC).
In GUT models, the quarks and leptons are placed in the same multiplet of the GUT group.
The massive gauge bosons which correspond to the broken generators of the GUT group
which change quarks into leptons, and vice versa, are vector leptoquarks.
In ETC models, the technicolor interaction will bind the techniquarks and the technileptons into scalar or vector bound states. 
These leptoquark states couple to the ordinary quarks and leptons through ETC interactions.

The interactions of leptoquarks with ordinary matter 
can be described in a model-independent fashion 
by an effective low-energy Lagrangian as discussed in Ref.~\cite{leptoquarks}.
Assuming the fermionic content of the SM, the most general
dimensionless $SU(3)_C\times SU(2)_L\times U(1)_Y$ invariant couplings of
scalar and vector leptoquarks satisfying baryon and lepton number 
conservation is given by:
\begin{equation}
{\cal L} = {\cal L}_{F=2} + {\cal L}_{F=0}\;,
\end{equation}
where
\begin{eqnarray}
{\cal L}_{F=2} &=& 
\Bigl[  
 g_{1L}\,\overline{q_L^c} i\tau_2 \ell_L^{\phantom{c}}
+g_{1R}\,\overline{u_R^c} e_R^{\phantom{c}}
\Bigr] S_1
+\tilde{g}_{1R}\Bigl[\overline{d_R^c} e_R^{\phantom{c}} \Bigr] \tilde{S}_1
\nonumber\\
&& 
+g_{3L}\Bigl[\overline{q_L^c} i\tau_2 \vec{\tau} \ell_L^{\phantom{c}} \Bigr] \vec{S}_3 
\nonumber\\
&& 
+\Bigl[
 g_{2L}\,\overline{d_R^c} \gamma^\mu \ell_L^{\phantom{c}}
+g_{2R}\,\overline{q_L^c} \gamma^\mu e_R^{\phantom{c}}
\Bigr] V_{2\mu}
+\tilde{g}_{2L}
\Bigl[\overline{u_R^c} \gamma^\mu \ell_L^{\phantom{c}} \Bigr] \tilde{V}_{2\mu}
+h.c. \;,
\label{lqlag:f2}
\\  \vsk{0.3}
{\cal L}_{F=0}
&=&
\Bigl[h_{2L}\,\overline{u_R}\ell_L + h_{2R}\,\overline{q_L}i\tau_2 e_R
\Bigr] S_2
+\tilde{h}_{2L} \Bigl[ \overline{d_R} \ell_L \Bigr] \tilde{S}_2
\nonumber\\
&&
+\Bigl[h_{1L}\,\overline{q_L}\gamma^\mu \ell_L +
       h_{1R}\,\overline{d_R}\gamma^\mu e_R 
\Bigr] V_{1\mu}
+\tilde{h}_{1R}
\Bigl[ \overline{u_R}\gamma^\mu e_R^{\phantom{c}} \Bigr] \tilde{V}_{1\mu}
\nonumber\\
&&
+h_{3L} \Bigl[ \overline{q_L} \vec{\tau}\gamma^\mu \ell_L
\Bigr] \vec{V}_{3\mu} 
+h.c. \;.
\label{lqlag:f0}
\end{eqnarray}
Here, the scalar and vector leptoquark fields are denoted by $S$ and $V$, respectively,
their subscripts indicating the dimension of their $SU(2)_L$ representation.
The same index is attached to their respective coupling constants, the $g$'s and $h$'s,
with the extra subscript $L$ or $R$ indicating the chirality of the lepton involved in the interaction.
For simplicity, color, weak isospin, and generation indices have been suppressed.
The leptoquarks $S_1, \tilde{S}_1, \vec{S}_3, V_2, \tilde{V}_2$ carry 
fermion number $F=3B+L=-2$, while
the leptoquarks $S_2, \tilde{S}_2, V_1, \tilde{V}_1, \vec{V}_3$ have $F=0$.

Rewriting the fermion doublets and the leptoquark multiplets 
in terms of the individual component fields, Eqs.~({\ref{lqlag:f2}}) and ({\ref{lqlag:f0}}) 
are expanded as follows:
\begin{eqnarray}
{\cal L}_{F=2} &=& 
\Bigl[
g_{1L}(\overline{u_L^c}e_L^{\phantom{c}} - \overline{d_L^c}\nu_L^{\phantom{c}} )
+g_{1R}(\overline{u_R^c}e_R^{\phantom{c}})
\Bigr] S_1^0
+\tilde{g}_{1R}\Bigl[ \overline{d_R^c}e_R^{\phantom{c}}\Bigr] \tilde{S}_1^0
\nonumber\\
&&
+\Bigl[
 g_{2L}(\overline{d_R^c}\gamma^\mu e_L^{\phantom{c}})
+g_{2R}(\overline{d_L^c}\gamma^\mu e_R^{\phantom{c}}) 
\Bigr] V_{2\mu}^+
+\Bigl[
 g_{2L}(\overline{d_R^c}\gamma^\mu \nu_L^{\phantom{c}})
+g_{2R}(\overline{u_L^c}\gamma^\mu e_R^{\phantom{c}}) 
\Bigr] V_{2\mu}^-
\nonumber\\
&&
+\tilde{g}_{2L}\Bigl[
(\overline{u_R^c}\gamma^\mu e_L^{\phantom{c}}) \tilde{V}_{2\mu}^+
+ (\overline{u_R^c}\gamma^\mu \nu_L^{\phantom{c}}) \tilde{V}_{2\mu}^-
\Bigr]
\nonumber\\
&&
+g_{3L}\Bigl[-\sqrt{2}(\overline{d_L^c}e_L^{\phantom{c}}) S_3^+
             -(\overline{u_L^c}e_L+\overline{d_L^c}\nu_L^{\phantom{c}}) S_3^0
             +\sqrt{2}(\overline{u_L^c}\nu_L^{\phantom{c}}) S_3^- \Bigr]
+ h.c. \;,
\\  \vsk{0.3}
{\cal L}_{F=0}
&=&
\Bigl[
h_{2L}(\overline{u_R}e_L)+h_{2R}(\overline{u_L}e_R)
\Bigr] S_2^+
+\Bigl[
h_{2L}(\overline{u_R}\nu_L)-h_{2R}(\overline{d_L}e_R)
\Bigr] S_2^-
\nonumber\\
&&
+\tilde{h}_{2L} \Bigl[
(\overline{d_R}e_L)\tilde{S}_2^+ + (\overline{d_R}\nu_L)\tilde{S}_2^-
\Bigr] 
\nonumber\\
&&
+\Bigl[
h_{1L}(\overline{u_L}\gamma^\mu \nu_L +
       \overline{d_L}\gamma^\mu e_L)
+h_{1R}(\overline{d_R}\gamma^\mu e_R) \Bigr]V_{1\mu}^0
+\tilde{h}_{1R}\Big[\overline{u_R}\gamma^\mu e_R\Bigr] \tilde{V}_{1\mu}^0
\nonumber\\
&&
+h_{3L}\Bigl[\sqrt{2}(\overline{u_L}\gamma^\mu e_L) V_{3\mu}^+
             +(\overline{u_L}\gamma^\mu \nu_L 
               -\overline{d_L}\gamma^\mu e_L) V_{3\mu}^0
             +\sqrt{2}(\overline{d_L}\gamma^\mu \nu_L) V_{3\mu}^- \Bigr] 
+ h.c. \;.
\end{eqnarray}
%
%
Superscripts indicate the weak isospin of each field, not the electromagnetic charge.
For fields with subscript $1$, the superscript $0$ is redundant and may be dropped.
The quantum numbers and couplings of the various leptoquarks fields are summarized in 
Table~\ref{LQtable}. 
Note that the scalar $\tilde{S}_1$ and the vector $\tilde{V}_{1\mu}$
do not couple to the neutrinos, so they are irrelevant to our discussion and will not be considered further.
The isospin plus components of the remaining leptoquarks, namely 
$S_2^+$, $\tilde{S}_2^+$, $S_3^+$, $V_{2\mu}^+$, $\tilde{V}_{2\mu}^+$, and $V_{3\mu}^+$,
do not couple to the neutrinos either, but we will keep them in our Lagrangian since their 
coupling constants are common with the other components that do couple, and are important in understanding how the couplings are constrained by neutrinoless experiments.

\begin{center}
\begin{table}[tbp]
\begin{tabular}{|c|c||c|c|c|c|c||c||c|}
\hline
\multicolumn{2}{|c|}{\ Leptoquark\ \ }  &\ Spin\ \ & $\;\;F\;\;$ & $\,SU(3)_C\,$ 
& $\quad I_3\quad$ & $\quad Y\quad$ & $\;\;Q_{em}\;\;$ &\ Allowed Couplings\ \\
\hline\hline
$\;S_1\;$ 
& $\;S_1^0\;$
& $0$ 
& $-2$ 
& $\bar{3}$ 
& $\phantom{+}0$ 
& $\phantom{+}\frac{1}{3}$ 
& $\phantom{+}\frac{1}{3}$ 
& $\;g_{1L}(\overline{u_L^c}e_L^{\phantom{c}}-\overline{d_L^c}\nu_L^{\phantom{c}}),\,
g_{1R}(\overline{u^c_R}e_R^{\phantom{c}})\;$ \\
\hline
$\;\tilde{S}_1\;$ 
& $\;\tilde{S}_1^0\;$
& $0$ 
& $-2$ 
& $\bar{3}$ 
& $\phantom{+}0$ 
& $\phantom{+}\frac{4}{3}$ 
& $\phantom{+}\frac{4}{3}$ 
& $\tilde{g}_{1R}(\overline{d_R^c}e_R^{\phantom{c}})$ \\
\hline
$\;V_{2\mu}\;$
& $\;V_{2\mu}^+\;$ 
& $1$ 
& $-2$ 
& $\bar{3}$ 
& $+\frac{1}{2}$ 
& $\phantom{+}\frac{5}{6}$ 
& $\phantom{+}\frac{4}{3}$ 
& $g_{2L}(\overline{d_R^c}\gamma^\mu e_L^{\phantom{c}}),
\,g_{2R}(\overline{d_L^c}\gamma^\mu e_R^{\phantom{c}}) $ \\
& $\;V_{2\mu}^-\;$ 
& 
& 
& 
& $-\frac{1}{2}$ 
& 
& $\phantom{+}\frac{1}{3}$ 
& $g_{2L}(\overline{d_R^c}\gamma^\mu \nu_L^{\phantom{c}}),\,
g_{2R}(\overline{u_L^c}\gamma^\mu e_R^{\phantom{c}}) $ \\
\hline
$\tilde{V}_{2\mu}\;$
& $\tilde{V}_{2\mu}^+\;$ 
& $1$ 
& $-2$ 
& $\bar{3}$ 
& $+\frac{1}{2}$ 
& $-\frac{1}{6}$ 
& $\phantom{+}\frac{1}{3}$ 
&  $\tilde{g}_{2L}(\overline{u_R^c}\gamma^\mu e_L^{\phantom{c}})$\\
& $\;\tilde{V}_{2\mu}^-\;$ 
& 
& 
& 
& $-\frac{1}{2}$ 
& 
& $-\frac{2}{3}$ 
& $\tilde{g}_{2L}(\overline{u_R^c}\gamma^\mu \nu_L^{\phantom{c}})$  \\
\hline
$\;\vec{S}_3\;$
& $\;S_3^+\;$ 
& $0$ 
& $-2$ 
& $\bar{3}$ 
& $+1$ 
& $\phantom{+}\frac{1}{3}$ 
& $\phantom{+}\frac{4}{3}$ 
& $-\sqrt{2} g_{3L}(\overline{d_L^c}e_L^{\phantom{c}}) $ \\
& $\;S_3^0\;$ 
& 
& 
& 
& $\phantom{+}0$ 
& 
& $\phantom{+}\frac{1}{3}$ 
& $-g_{3L}(\overline{u_L^c}e_L^{\phantom{c}}+\overline{d_L^c}\nu_L^{\phantom{c}}) $ \\
& $\;S_3^-\;$ 
& 
& 
& 
& $-1$ 
&
& $-\frac{2}{3}$
& $\sqrt{2}g_{3L}(\overline{u_L^c}\nu_L^{\phantom{c}})$ \\
\hline\hline
$\;S_2\;$
& $\;S_2^+\;$ 
& $0$ 
& $0$ 
& $3$ 
& $+\frac{1}{2}$ 
& $\phantom{+}\frac{7}{6}$ 
& $\phantom{+}\frac{5}{3}$ 
&  $h_{2L}(\overline{u_R} e_L),h_{2R}(\overline{u_L}e_R) $ \\
& $\;S_2^-\;$ 
&
&
&
& $-\frac{1}{2}$ 
& 
& $\phantom{+}\frac{2}{3}$ 
& $h_{2L}(\overline{u_R} \nu_L),-h_{2R}(\overline{d_L}e_R) $ \\
\hline
$\;\tilde{S}_2\;$ 
& $\;\tilde{S}_2^+\;$ 
& $0$ 
& $0$ 
& $3$ 
& $+\frac{1}{2}$ 
& $\phantom{+}\frac{1}{6}$ 
& $\phantom{+}\frac{2}{3}$ 
& $\tilde{h}_{2L}(\overline{d_R}e_L)$ \\
& $\;\tilde{S}_2^-\;$ 
&
&
&
& $-\frac{1}{2}$ 
&
& $-\frac{1}{3}$ 
& $\tilde{h}_{2L}(\overline{d_R}\nu_L)$ \\
\hline
$\;V_{1\mu}\;$
& $\;V_{1\mu}^0\;$ 
& $1$ 
& $0$ 
& $3$ 
& $\phantom{+}0$ 
& $\phantom{+}\frac{2}{3}$ 
& $\phantom{+}\frac{2}{3}$ 
& $\;h_{1L}(\overline{u_L}\gamma^\mu \nu_L+\overline{d_L}\gamma^\mu e_L),\;h_{1R}(\overline{d_R}\gamma^\mu e_R)\;$ \\
\hline
$\;\tilde{V}_{1\mu}\;$
& $\;\tilde{V}_{1\mu}^0$ 
& $1$ 
& $0$ 
& $3$ 
& $\phantom{+}0$ 
& $\phantom{+}\frac{5}{3}$ 
& $\phantom{+}\frac{5}{3}$ 
& $\tilde{h}_{1R}(\overline{u_R}\gamma^\mu e_R)$ \\
\hline
$\;\vec{V}_{3\mu}\;$
& $\;V_{3\mu}^+\;$ 
& $1$ 
& $0$ 
& $3$ 
& $+1$ 
& $\phantom{+}\frac{2}{3}$ 
& $\phantom{+}\frac{5}{3}$ 
& $\sqrt{2}h_{3L}(\overline{u_L}\gamma^\mu e_L)$ \\
& $\;V_{3\mu}^0\;$ 
&
&
&
& $\phantom{+}0$ 
&  
& $\phantom{+}\frac{2}{3}$ 
& $h_{3L}(\overline{u_L}\gamma^\mu \nu_L-\overline{d_L}\gamma^\mu e_L) $ \\
& $V_{3\mu}^-\;$ 
&
& 
&
& $-1$ 
&  
& $-\frac{1}{3}$ 
& $\sqrt{2}h_{3L}(\overline{d_L}\gamma^\mu \nu_L)$ \\
\hline
\end{tabular}
\caption[]{Quantum numbers of scalar and vector leptoquarks with
$SU(3)_C\times SU(2)_L\times U(1)_Y$ invariant couplings to quark-lepton
pairs ($Q_{\rm em}=I_3+Y$).}
\label{LQtable}
\end{table}
\end{center}

Since the leptoquarks must distinguish among different generation fermions
to contribute to neutrino oscillation matter effects, 
we generalize their interactions by allowing the
coupling constants to depend on the generations of the quarks and leptons that couple
to each leptoquark:
%
\begin{eqnarray}
{\cal L}_{F=2} &=& 
\Bigl[
g_{1L}^{ij}(\overline{u_{iL}^c}e_{jL}^{\phantom{c}} - \overline{d_{iL}^c}\nu_{jL}^{\phantom{c}} )
+g_{1R}^{ij}(\overline{u_{iR}^c}e_{jR}^{\phantom{c}})
\Bigr] S_{1}^0
\nonumber\\
&&
+\Bigl[
g_{2L}^{ij}(\overline{d_{iR}^c}\gamma^\mu e_{jL}^{\phantom{c}})
+g_{2R}^{ij}(\overline{d_{iL}^c}\gamma^\mu e_{jR}^{\phantom{c}}) 
\Bigr] V_{2\mu}^+
+\Bigl[
 g_{2L}^{ij}(\overline{d_{iR}^c}\gamma^\mu \nu_{jL}^{\phantom{c}})
+g_{2R}^{ij}(\overline{u_{iL}^c}\gamma^\mu e_{jR}^{\phantom{c}}) 
\Bigr] V_{2\mu}^-
\nonumber\\
&&
+\tilde{g}_{2L}^{ij}\Bigl[
(\overline{u_{iR}^c}\gamma^\mu e_{jL}^{\phantom{c}}) \tilde{V}_{2\mu}^+
+ (\overline{u_{iR}^c}\gamma^\mu \nu_{jL}^{\phantom{c}}) \tilde{V}_{2\mu}^-
\Bigr]
\nonumber\\
&&
+g_{3L}^{ij}\Bigl[-\sqrt{2}(\overline{d_{iL}^c}e_{jL}^{\phantom{c}}) S_{3}^+
             -(\overline{u_{iL}^c}e_{jL}^{\phantom{c}}+\overline{d_{iL}^c}\nu_{jL}^{\phantom{c}}) S_{3}^0
             +\sqrt{2}(\overline{u_{iL}^c}\nu_{jL}^{\phantom{c}}) S_{3}^- \Bigr] + h.c.\;,
\\  \vsk{0.3}
{\cal L}_{F=0}
&=&
\Bigl[
h_{2L}^{ij}(\overline{u_{iR}}e_{jL})+h_{2R}^{ij}(\overline{u_{iL}}e_{jR})
\Bigr] S_{2}^+
+\Bigl[
h_{2L}^{ij}(\overline{u_{iR}}\nu_{jL})-h_{2R}^{ij}(\overline{d_{iL}}e_{jR})
\Bigr] S_{2}^-
\nonumber\\
&&
+\tilde{h}_{2L}^{ij} \Bigl[
(\overline{d_{iR}}e_{jL})\tilde{S}_{2}^+ + (\overline{d_{iR}}\nu_{jL})\tilde{S}_{2}^-
\Bigr] 
\nonumber\\
&&
+\Bigl[
h_{1L}^{ij}(\overline{u_{iL}}\gamma^\mu \nu_{jL} +
       \overline{d_{iL}}\gamma^\mu e_{jL})
+h_{1R}^{ij}(\overline{d_{iR}}\gamma^\mu e_{jR}) \Bigr]V_{1\mu}^0
\nonumber\\
&&
+h_{3L}^{ij}\Bigl[\sqrt{2}(\overline{u_{iL}}\gamma^\mu e_{jL}) V_{3\mu}^+
             +(\overline{u_{iL}}\gamma^\mu \nu_{jL} 
               -\overline{d_{iL}}\gamma^\mu e_{jL}) V_{3\mu}^0
             +\sqrt{2}(\overline{d_{iL}}\gamma^\mu \nu_{jL}) V_{3\mu}^- \Bigr]
+ h.c. \;.\cr
& &
\end{eqnarray}
%
Here, $i$ is the quark generation number, and $j$ is the lepton generation number. 
Summation over repeated indices is assumed. 
The interactions that contribute to neutrino oscillation matter effects are 
those with indices $(ij)=(12)$ and $(ij)=(13)$.
It is often assumed in the literature that generation non-diagonal
couplings are absent to account for the non-observation 
of flavor changing neutral currents and lepton flavor violation. 
However, the constraints from such rare processes are always on 
\textit{products of different $(ij)$-couplings} and not on the \textit{individual} non-diagonal couplings by themselves.  For instance,
non-observation of the decay $K_L\rightarrow \bar{e}\mu$ constrains
the product of $(12)$ and $(21)$ couplings, but not the $(12)$ and $(21)$ 
couplings separately, which allows one of them to be sizable if the other is small.
Constraints on the individual $(12)$ and $(13)$ couplings actually come from 
precision measurements of \textit{flavor conserving} processes, such as
$R_\pi = \Gamma(\pi\rightarrow \mu\nu_\mu)/\Gamma(\pi\rightarrow e\nu_e)$
which constrains the square of the $(12)$ coupling, 
and those constraints are not yet that strong \cite{davidson}.

In the following, we calculate the effective value of $\xi$ induced by the exchange of these leptoquarks.
The leptoquark fields are naturally grouped into pairs from the way they couple to the
quarks and leptons: $(S_1,\vec{S}_3)$, $(S_2,\tilde{S}_2)$, $(V_2,\tilde{V}_2)$,
and $(V_1,\vec{V}_3)$. We treat each of these pairs in turn, 
and then discuss the potential bounds on the leptoquark couplings and masses.


\subsection{$S_1$ and $\vec{S}_3$ leptoquarks}

\begin{figure}[ht]
\centering
    \begin{picture}(400,120)(-100,-80) 
    \SetWidth{1}
    \SetScale{1}  
    \SetColor{Black}
    
    \ArrowLine(-80,-40)(-40,0)
    \ArrowLine(-80,40)(-40,0)
    \DashArrowLine(40,0)(-40,0){5}
    \ArrowLine(40,0)(80,40)
    \ArrowLine(40,0)(80,-40)
    \Vertex(-40,0){2}
    \Vertex(40,0){2}       
    \Text(-70,50)[]{$\nu_j(k)$}
    \Text(70,50)[]{$d(p)$}
    \Text(-70,-50)[]{$d(p)$}
    \Text(70,-50)[]{$\nu_j(k)$}
    \Text(0,15)[]{$S_{\alpha}^{0}$}
    \Text(0,-20)[]{$p+k$}
    \Text(-70,0)[]{$-i\, g_{\alpha L}^{1j}$}
    \Text(70,0)[]{$-i\, g_{\alpha L}^{1j*}$}
    \Text(0,-75)[]{$(a)\;\alpha=1, 3$}
    \SetWidth{0.5}
    \LongArrow(-15,-10)(15,-10)

    \SetWidth{1.0}
    \SetOffset(200,0)	
    \ArrowLine(-80,-40)(-40,0)
    \ArrowLine(-80,40)(-40,0)
    \DashArrowLine(40,0)(-40,0){5}
    \ArrowLine(40,0)(80,40)
    \ArrowLine(40,0)(80,-40)
    \Vertex(-40,0){2}
    \Vertex(40,0){2}       
    \Text(-70,50)[]{$\nu_j(k)$}
    \Text(70,50)[]{$u(p)$}
    \Text(-70,-50)[]{$u(p)$}
    \Text(70,-50)[]{$\nu_j(k)$}
    \Text(0,15)[]{$S_{3}^{-}$}
    \Text(0,-20)[]{$p+k$}
    \Text(-70,0)[]{$i\sqrt{2}\, g_{3L}^{1j}$}
    \Text(70,0)[]{$i\sqrt{2}\, g_{3L}^{1j*}$}
    \Text(0,-75)[]{$(b)$}    
    \SetWidth{0.5}
    \LongArrow(-15,-10)(15,-10)

    \end{picture}
	\caption{
	Diagrams contributing to neutrino oscillation matter effects from the exchange of (a) $S_1^0$
	or the isospin $0$ component of $\vec{S}_3$, and (b) the isospin $-1$ component of $\vec{S}_3$. 
	The EM charge $Q_{em}=I_3+Y$ for $S_1^0$ and $S_3^0$ are $+\frac{1}{3}$, 
	while that for $S_3^{-}$ is $-\frac{2}{3}$. 
	}
	\label{LQ12}
\end{figure}
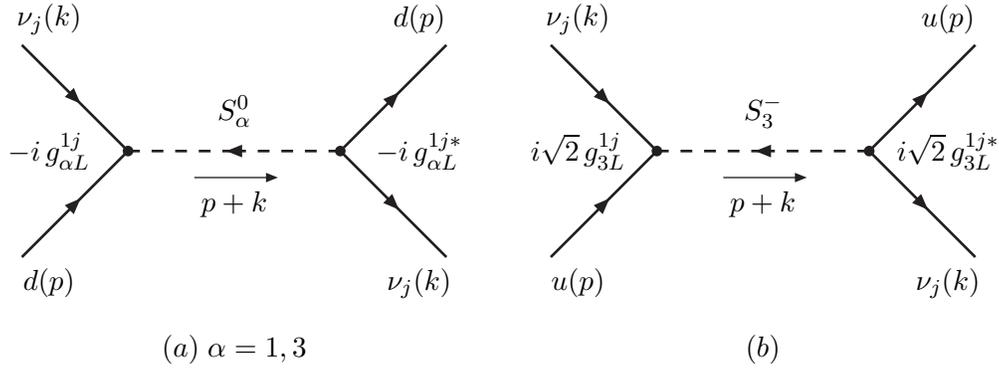

The $(ij)=(12)$ and $(13)$ interactions of the leptoquarks $S_1$ and $\vec{S}_3$ are,
respectively,
\begin{equation} \label{S1}
\mathcal{L} =
- g_{1L}^{12} (\overline{d_{L}^c}\nu_{\mu L}^{\phantom{c}}) S_{1}
- g_{1L}^{13} (\overline{d_{L}^c}\nu_{\tau L}^{\phantom{c}}) S_{1}
+ h.c. \;,
\end{equation}
and 
\begin{eqnarray}\label{vec S3}
\mathcal{L} & = & 
  g_{3L}^{12}\left[ -(\overline{d_{L}^c}\nu_{\mu L}) S_{3}^0
          +\sqrt{2}(\overline{u_{L}^c}\nu_{\mu L}) S_{3}^-
           \right]
+ g_{3L}^{13}\left[ -(\overline{d_{L}^c}\nu_{\tau L}) S_{3}^0
          +\sqrt{2}(\overline{u_{L}^c}\nu_{\tau L}) S_{3}^-
           \right]
+ h.c. 
\cr & &
\end{eqnarray}
The interactions described by Eqs.~(\ref{S1}) and (\ref{vec S3}) 
can be written in a common general form as
\begin{equation}
\mathcal{L} = 
\lambda\,(\overline{q^c}P_L\,\nu) S+\lambda^*(\overline{\nu}P_R\,q^c) \bar{S}\;,
\end{equation}
where $q=u$ or $d$.
The Feynman diagrams contributing to neutrino oscillation matter effects are shown in Fig.~\ref{LQ12}. 
At momenta much smaller than the mass of the leptoquark, the corresponding matrix element is
\begin{equation} \label{me1}
i\mathcal{M}=(-i)^2|\lambda|^2
\bra{\nu, q} \left( \overline\nu P_R\,q^c \right) \left(\frac{-i}{M^2_{S}}\right)
\left( \overline{q^c}P_L\,\nu \right) \ket{\nu, q}\;.
\end{equation}
Using the Fiertz rearrangement
\begin{equation}  \label{Fiertz1}
\left( \overline\nu P_R\, q^c \right) \left( \overline{q^c}P_L\,\nu \right) 
\;=\; -\frac{1}{2}\left( \overline{\nu} \gamma^\mu P_L\,\nu \right) \left( \overline{q^c}\gamma_\mu P_R\,q^c \right)
\;=\; +\frac{1}{2}\left( \overline{\nu} \gamma^\mu P_L\,\nu \right) \left( \overline{q}\gamma_\mu P_L\,q \right)\;,
\end{equation}
we obtain
\begin{equation}
i\mathcal{M}=\frac{i|\lambda|^2}{2M^2_{S}}\,
\bra{\nu} \overline \nu\gamma^{\mu}P_L\nu \ket{\nu}\,
\bra{q} \overline{q} \gamma_{\mu}P_L q \ket{q}
\rightarrow i\frac{|\lambda|^2}{4 M_{S}^2}\,N_q \left(\phi_{\nu}^{\dagger}\phi_{\nu}^{\phantom{\dagger}}\right)
=-iV_{\nu}\left(\phi_{\nu}^{\dagger}\phi_{\nu}^{\phantom{\dagger}}\right)\;,
\end{equation}
where
\begin{equation} 
V_{\nu}
\equiv -\frac{N_q}{4} \frac{|\lambda|^2}{M^2_{S}}\;.
\label{potential-0}
\end{equation}
Applying this expression to the $S_1$ case, the effective potential
for the neutrino of generation number $j$ is:
\begin{equation}
V_{\nu_j} 
\;=\; - \dfrac{N_d}{4}\dfrac{\left|g_{1L}^{1j}\right|^2}{M_{S_{1}}^2} 
\;=\; - \dfrac{(N_p+2N_n)}{4}\dfrac{\left|g_{1L}^{1j}\right|^2}{M_{S_{1}}^2}
\;\approx\; -\dfrac{3N}{4}\dfrac{\left|g_{1L}^{1j}\right|^2}{M_{S_{1}}^2}\;,
\end{equation}
The effective $\xi$ is then
\begin{equation}
\xi_{S_{1}}
\;=\; \dfrac{V_{\nu_3}-V_{\nu_2}}{V_{CC}}
\;=\; +3\;
\dfrac{(\;\left| g_{1L}^{12} \right|^2 - \left| g_{1L}^{13} \right|^2\;) / M_{S_{1}}^2}
          {g^2/M_W^2} \;.
\end{equation}
For the $\vec{S}_3$ case, the effective potential is
\begin{eqnarray}
V_{\nu_j} & = & 
-\dfrac{N_d}{4}\dfrac{|g_{3L}^{1j}|^2}{M_{S_3^0}^2}
-\dfrac{N_u}{2}\dfrac{|g_{3L}^{1j}|^2}{M_{S_3^-}^2} \cr
& = & -|g_{3L}^{1j}|^2
\left[\dfrac{(N_p+2N_n)}{4M_{S_3^0}^2}-\dfrac{(2N_p+N_n)}{2M_{S_3^-}^2}\right] \cr
& \approx &
-\dfrac{3N}{4}
\left|g_{3L}^{1j}\right|^2
\left(
\dfrac{1}{M_{S_{3}^0}^2}+\dfrac{2}{M_{S_{3}^-}^2}
\right) \;,
\end{eqnarray}
and the effective $\xi$ is
\begin{eqnarray}
\xi_{\vec{S}_3}
\;=\; \dfrac{V_{\nu_3}-V_{\nu_2}}{V_{CC}}
\;=\; +3\;
\frac{\;\left| g_{3L}^{12} \right|^2 - \left| g_{3L}^{13} \right|^2\;}{g^2/M_W^2}
\left( \dfrac{1}{M_{S_{3}^0}^2}+\dfrac{2}{M_{S_{3}^-}^2} \right) \;.
\end{eqnarray}
In the case of degenerate mass, $M_{S_3^0} = M_{S_3^-} \equiv M_{S_3}$,
we have
\begin{equation}
\xi_{\vec{S}_3}
\;=\; +9\;
\dfrac{(\;\left| g_{3L}^{12} \right|^2 - \left| g_{3L}^{13} \right|^2 \;) / M_{S_3}^2 }
          {g^2/M_W^2} \;.
\end{equation}


\subsection{$S_2$ and $\tilde{S}_2$ leptoquarks}

\begin{figure}[t]
\centering
    \begin{picture}(400,130)(-100,-90) 
    \SetWidth{1}
    \SetScale{1}  
    \SetColor{Black}
    
    \ArrowLine(-80,-40)(0,-40)
    \ArrowLine(0,-40)(80,-40)
    \DashArrowLine(0,-40)(0,40){5}
    \ArrowLine(-80,40)(0,40)
    \ArrowLine(0,40)(80,40)
    \Vertex(0,40){2}
    \Vertex(0,-40){2}       
    \Text(-70,50)[]{$\nu_j(k)$}
    \Text(70,50)[]{$u(p)$}
    \Text(-70,-50)[]{$u(p)$}
    \Text(70,-50)[]{$\nu_j(k)$}
    \Text(-15,0)[]{$S_2^-$}
    \Text(25,0)[]{$k-p$}
    \Text(0,53)[]{$+i\, h_{2L}^{1j}$}
    \Text(0,-53)[]{$+i\, h_{2L}^{1j}$}
    \Text(0,-85)[]{$(a)$}
    \SetWidth{0.5}
    \LongArrow(7,15)(7,-15)

    \SetOffset(200,0)
    \SetWidth{1.0}
    \ArrowLine(-80,-40)(0,-40)
    \ArrowLine(0,-40)(80,-40)
    \DashArrowLine(0,-40)(0,40){5}
    \ArrowLine(-80,40)(0,40)
    \ArrowLine(0,40)(80,40)
    \Vertex(0,40){2}
    \Vertex(0,-40){2}       
    \Text(-70,50)[]{$\nu_j(k)$}
    \Text(70,50)[]{$d(p)$}
    \Text(-70,-50)[]{$d(p)$}
    \Text(70,-50)[]{$\nu_j(k)$}
    \Text(-15,0)[]{$\tilde{S}_2^-$}
    \Text(25,0)[]{$k-p$}
    \Text(0,53)[]{$+i\,\tilde{h}_{2L}^{1j}$}
    \Text(0,-53)[]{$+i\,\tilde{h}_{2L}^{1j}$}
    \Text(0,-85)[]{$(b)$}
    \SetWidth{0.5}
    \LongArrow(7,15)(7,-15)
    
    \end{picture}
	\caption{
	Diagrams contributing to neutrino oscillation matter effects from the exchange  of
	(a) $S_2^{-}$, and (b) $\tilde{S}_2^{-}$.
	The EM charge $Q_{em}=I_3+Y$ for $S_2^-$ is $+\frac{2}{3}$, 
	while that for $\tilde{S}_2^{-}$ is $-\frac{1}{3}$. 
	}
	\label{LQ34}
\end{figure}
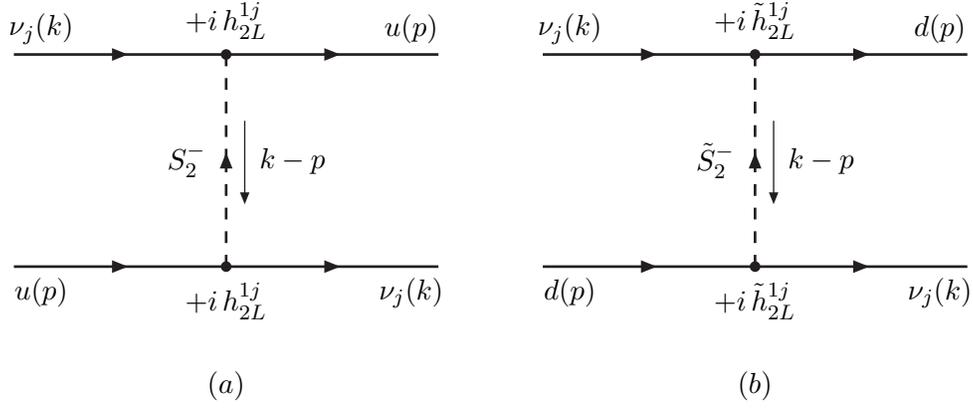

The relevant interactions are
\begin{equation} \label{S2}
\mathcal{L} =
h_{2L}^{12} (\overline{u_{R}}\nu_{\mu L}) S_{2}^-
+ h_{2L}^{13} (\overline{u_{R}}\nu_{\tau L}) S_{2}^-
+ h.c.
\end{equation}
for $S_2^-$ and 
\begin{equation} \label{tilde S2}
\mathcal{L} =
\tilde{h}_{2L}^{12} (\overline{d_{R}}\nu_{\mu L})\tilde{S}_{2}^-
+ \tilde{h}_{2L}^{13} (\overline{d_{R}}\nu_{\tau L})\tilde{S}_{2}^-
+ h.c.
\end{equation}
for $\tilde S_2^-$ leptoquarks. Both (\ref{S2}) and (\ref{tilde S2}) can be written in a common general form as
\begin{equation} 
\mathcal{L} 
= \lambda\,(\overline qP_L\,\nu) S + \lambda^*(\overline{\nu} P_R\,q)\bar{S}\;,
\end{equation}
where $q=u$ or $d$.
The Feynman diagram contributing to neutrino oscillation matter effects is shown in Fig.~\ref{LQ34}a. For momenta much smaller than the mass of the leptoquark, the corresponding matrix element is
\begin{equation}
i\mathcal{M}  = (-i)^2 |\lambda|^2
\bra{\nu,q} \left(\overline{\nu}P_R\, q\right) \left(\frac{\displaystyle -i}{\displaystyle M^2_{S}}\right)
\left( \overline{q} P_L\, \nu \right) \ket{\nu,q} \;.
\end{equation}
Using the Fiertz identity given in Eq.~(\ref{Fiertz1}) again, we obtain
\begin{equation}
i\mathcal{M}  
= -i\frac{\displaystyle |\lambda|^2}{\displaystyle 2 M^2_{S}}
\,\bra{\nu} \overline{\nu}\gamma^{\mu}P_L\,\nu\ket{\nu}
\,\bra{q} \overline{q}\gamma_{\mu}P_R\,q \ket{q} 
\rightarrow -i\frac{|\lambda|^2}{4 M_{S}^2}N_q
\left( \phi_{\nu}^{\dagger}\phi_{\nu}^{\phantom{\dagger}} \right) =
-iV_{\nu} \left( \phi_{\nu}^{\dagger}\phi_{\nu}^{\phantom{\dagger}} \right) \;,
\end{equation}
where
\begin{equation} \label{potential-1}
V_{\nu}
\;=\; +\frac{N_q}{4}\frac{|\lambda|^2}{M^2_{S}}\;.
\end{equation}
Applying this expression to the $S_2^-$ case, the effective potential
for the neutrino of generation number $j$ is
\begin{equation}
V_{\nu_j} 
\;=\; +\frac{N_u}{4}\dfrac{\left|h_{2L}^{1j}\right|^2}{M_{S_{2}^-}^2} 
\;=\; +\frac{(2N_p+N_n)}{4}\dfrac{\left|h_{2L}^{1j}\right|^2}{M_{S_{2}^-}^2} 
\;\approx\; +\frac{3N}{4}\dfrac{\left|h_{2L}^{1j}\right|^2}{M_{S_{2}^-}^2}
\;,
\end{equation}
and the effective $\xi$ is
\begin{equation}
\xi_{S_2^-}
\;=\; \dfrac{V_{\nu_3}-V_{\nu_2}}{V_{CC}}
\;=\; -3\;
\dfrac{(\; \left| h_{2L}^{12} \right|^2 - \left| h_{2L}^{13} \right|^2 \;) /M_{S_{2}^-}^2}
          {g^2/M_W^2} \;.
\end{equation}
The effective potential for the $\tilde{S}_2^-$ case is
\begin{equation}
V_{\nu_j} 
\;=\; +\frac{N_d}{4}\dfrac{|\tilde{h}{}_{2L}^{1j}|^2}{M_{\tilde{S}_{2}^-}^2} 
\;=\; +\frac{(N_p+2N_n)}{4}\dfrac{|\tilde{h}{}_{2L}^{1j}|^2}{M_{\tilde{S}_{2}^-}^2} 
\;\approx\; +\frac{3N}{4}\dfrac{|\tilde{h}{}_{2L}^{1j}|^2}{M_{\tilde{S}_{2}^-}^2}
\;,
\end{equation}
and the effective $\xi$ is
\begin{equation}
\xi_{\tilde S_2^-}
\;=\; \dfrac{V_{\nu_3}-V_{\nu_2}}{V_{CC}}
\;=\; -3\;
\dfrac{(\; | \tilde{h}{}_{2L}^{12} |^2 - | \tilde{h}{}_{2L}^{13} |^2 \;) / M_{\tilde S_{2}^-}^2}
             {g^2/M_W^2} \;.
\end{equation}


\subsection{$V_{2}$ and $\tilde V_2$}


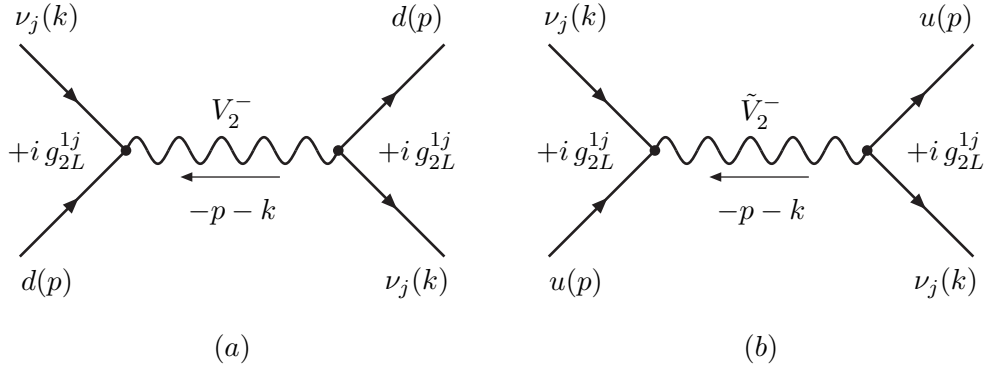
\begin{figure}[ht]
\centering
    \begin{picture}(400,120)(-100,-80) 
    \SetScale{1}  
    \SetColor{Black}
    
    \SetWidth{1}
    \ArrowLine(-80,-40)(-40,0)
    \ArrowLine(-80,40)(-40,0)
    \Photon(-40,0)(40,0){5}{5}
    \ArrowLine(40,0)(80,40)
    \ArrowLine(40,0)(80,-40)
    \Vertex(-40,0){2}
    \Vertex(40,0){2}       
    \Text(-70,50)[]{$\nu_j(k)$}
    \Text(70,50)[]{$d(p)$}
    \Text(-70,-50)[]{$d(p)$}
    \Text(70,-50)[]{$\nu_j(k)$}
    \Text(0,15)[]{$V_{2}^-$}
    \Text(0,-23)[]{$-p-k$}
    \Text(-70,0)[]{$+i\, g_{2L}^{1j}$}
    \Text(70,0)[]{$+i\, g_{2L}^{1j}$}
    \Text(0,-75)[]{$(a)$}
    \SetWidth{0.5}
    \LongArrow(18,-10)(-18,-10)

    \SetOffset(200,0)
    \SetWidth{1}
    \ArrowLine(-80,-40)(-40,0)
    \ArrowLine(-80,40)(-40,0)
    \Photon(-40,0)(40,0){5}{5}
    \ArrowLine(40,0)(80,40)
    \ArrowLine(40,0)(80,-40)
    \Vertex(-40,0){2}
    \Vertex(40,0){2}       
    \Text(-70,50)[]{$\nu_j(k)$}
    \Text(70,50)[]{$u(p)$}
    \Text(-70,-50)[]{$u(p)$}
    \Text(70,-50)[]{$\nu_j(k)$}
    \Text(0,15)[]{$\tilde{V}_{2}^-$}
    \Text(0,-23)[]{$-p-k$}
    \Text(-70,0)[]{$+i\, g_{2L}^{1j}$}
    \Text(70,0)[]{$+i\, g_{2L}^{1j}$}
    \Text(0,-75)[]{$(b)$}    
    \SetWidth{0.5}
    \LongArrow(18,-10)(-18,-10)
    \end{picture}
	\caption{
	Diagrams contributing to neutrino oscillation matter effects from the exchange of  
	(a) $V_2^{-}$, and (b) $\tilde{V}_2^{-}$.
	The EM charge $Q_{em}=I_3+Y$ for $V_2^-$ is $+\frac{1}{3}$, 
	while that for $\tilde{V}_2^{-}$ is $-\frac{2}{3}$. 
	}
	\label{LQ56}
\end{figure}


The relevant interactions for $V_2^-$ are
\begin{equation} \label{v1}
\mathcal{L} =
  g_{2L}^{12}(\overline{d_{R}^c}\gamma^\mu \nu_{\mu L})V_{2\mu}^-
+ g_{2L}^{13}(\overline{d_{R}^c}\gamma^\mu \nu_{\tau L})V_{2\mu}^-
+ h.c.
\end{equation}
and those for $\tilde{V}_2^-$ are 
\begin{equation} \label{v2}
\mathcal{L} =
  \tilde{g}_{2L}^{12}(\overline{u_{R}^c}\gamma^\mu \nu_{\mu L})\tilde{V}_{2\mu}^-
+ \tilde{g}_{2L}^{13}(\overline{u_{R}^c}\gamma^\mu \nu_{\tau L})\tilde{V}_{2\mu}^-
+ h.c.
\end{equation}
Both (\ref{v1}) and (\ref{v2}) can be written in a common general form as
\begin{equation} 
\mathcal{L} = \lambda\,(\overline{q^c}\gamma^{\mu} P_L\, \nu) V_{\mu}
+\lambda^*(\overline{\nu}\gamma^{\mu} P_L\, q^c) \bar{V}_{\mu} \;.
\end{equation}
The Feynman diagrams contributing to neutrino oscillation matter effects are shown in Fig.~\ref{LQ56}. For momenta much smaller than the mass of the leptoquark the corresponding matrix element is
\begin{equation}
i\mathcal{M}  =  (-i)^2|\lambda|^2
\bra{\nu,q} \left( \overline{\nu}\gamma^{\mu}P_L\, q^c \right)
\left( \frac{\displaystyle i}{\displaystyle M^2_{V}} \right)
\left( \overline{q^c}\gamma_{\mu} P_L\, \nu \right) \ket{\nu,q}\;.
\end{equation}
Using the Fiertz rearrangement
\begin{equation}  \label{Fiertz2}
\left( \overline\nu \gamma^\mu P_L\, q^c \right) \left( \overline{q^c}\gamma_\mu P_L\,\nu \right) 
\;=\; \left( \overline{\nu} \gamma^\mu P_L\,\nu \right) \left( \overline{q^c}\gamma_\mu P_L\,q^c \right)
\;=\; -\left( \overline{\nu} \gamma^\mu P_L\,\nu \right) \left( \overline{q}\gamma_\mu P_R\,q \right)\;,
\end{equation}
we obtain
\begin{equation}
i\mathcal{M}  =  i\frac{\displaystyle |\lambda|^2}{\displaystyle M^2_{V}}
\,\bra{\nu} \overline\nu\gamma^{\mu}P_L\,\nu \ket{\nu}
\,\bra{q} \overline q\gamma_{\mu}P_R\, q \ket{q}
\rightarrow i\frac{|\lambda|^2}{2 M_{V}^2}N_q \left( \phi_{\nu}^{\dagger}\phi_{\nu}^{\phantom{\dagger}} \right)
= -iV_{\nu} \left( \phi_{\nu}^{\dagger}\phi_{\nu}^{\phantom{\dagger}} \right)\;,
\end{equation}
where
\begin{equation} \label{potential-3}
V_{\nu}\equiv -\frac{N_q}{2}\frac{|\lambda|^2}{M^2_{V}}\;.
\end{equation}
Applying this to the $V_2^-$ case, the effective potential for
the neutrino of generation number $j$ is
\begin{equation}
V_{\nu_j} 
\;=\; - \dfrac{N_d}{2}\dfrac{\left|g_{2L}^{1j}\right|^2}{M_{V_{2}^-}^2}
\;=\; - \dfrac{(N_p+2N_n)}{2}\dfrac{\left|g_{2L}^{1j}\right|^2}{M_{V_{2}^-}^2}
\;\approx\; -  \dfrac{3N}{2}\dfrac{\left|g_{2L}^{1j}\right|^2}{M_{V_{2}^-}^2}\;.
\end{equation}
The effective $\xi$ is
\begin{equation}
\xi_{V_2^-}
\;=\; \dfrac{V_{\nu_3}-V_{\nu_2}}{V_{CC}}
\;=\; +6\;
\dfrac{(\; \left| g_{2L}^{12} \right|^2 - \left| g_{2L}^{13} \right|^2 \;) / M_{V_{2}^-}^2}
          {g^2/M_W^2} \;.
\end{equation}
The effective potential for the $\tilde{V}_{2}^-$ case is
\begin{equation}
V_{\nu_j} 
\;=\; -\dfrac{N_u}{2}\dfrac{\left|\tilde{g}{}_{2L}^{12}\right|^2}{M_{\tilde{V}_{2}^-}^2} 
\;=\; -\dfrac{(2N_p+N_n)}{2}\dfrac{\left|\tilde{g}{}_{2L}^{12}\right|^2}{M_{\tilde{V}_{2}^-}^2} 
\;\approx\; -\dfrac{N_u}{2}\dfrac{\left|\tilde{g}{}_{2L}^{12}\right|^2}{M_{\tilde{V}_{2}^-}^2} \;.
\end{equation}
The effective $\xi$ is
\begin{equation}
\xi_{\tilde{V}_{2}^-}
\;=\; \dfrac{V_{\nu_3}-V_{\nu_2}}{V_{CC}}
\;=\; +6\;
\dfrac{(\; \left| \tilde{g}{}_{2L}^{12} \right|^2 - \left| \tilde{g}{}_{2L}^{13} \right|^2 \;) / M_{\tilde V_{2}^-}^2}
          {g^2/M_W^2} \;.
\end{equation}
%


\subsection{$V_{1}$ and $\vec V_3$ leptoquarks}

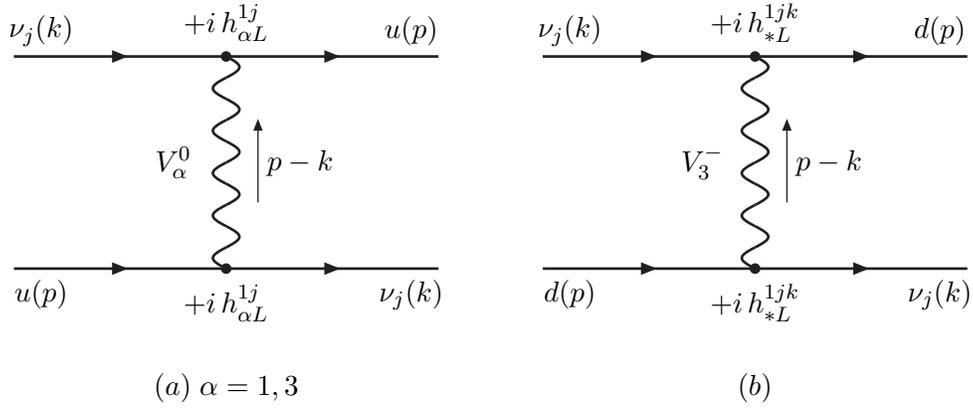
\begin{figure}[ht]
\centering
    \begin{picture}(400,130)(-100,-90) 
    \SetScale{1}  
    \SetColor{Black}
    
    \SetWidth{1}
    \ArrowLine(-80,-40)(0,-40)
    \ArrowLine(0,-40)(80,-40)
    \Photon(0,-40)(0,40){5}{5}
    \ArrowLine(-80,40)(0,40)
    \ArrowLine(0,40)(80,40)
    \Vertex(0,40){2}
    \Vertex(0,-40){2}       
    \Text(-70,50)[]{$\nu_j(k)$}
    \Text(70,50)[]{$u(p)$}
    \Text(-70,-50)[]{$u(p)$}
    \Text(70,-50)[]{$\nu_j(k)$}
    \Text(-20,0)[]{$V_{\alpha}^0$}
    \Text(28,0)[]{$p-k$}
    \Text(0,53)[]{$+i\,h_{\alpha L}^{1j}$}
    \Text(0,-53)[]{$+i\,h_{\alpha L}^{1j}$}
    \Text(0,-85)[]{$(a)\;\alpha=1,3$}
    \SetWidth{0.5}
    \LongArrow(12,-15)(12,15)

    \SetOffset(200,0)    
    \SetWidth{1}
    \ArrowLine(-80,-40)(0,-40)
    \ArrowLine(0,-40)(80,-40)
    \Photon(0,-40)(0,40){5}{5}
    \ArrowLine(-80,40)(0,40)
    \ArrowLine(0,40)(80,40)
    \Vertex(0,40){2}
    \Vertex(0,-40){2}       
    \Text(-70,50)[]{$\nu_j(k)$}
    \Text(70,50)[]{$d(p)$}
    \Text(-70,-50)[]{$d(p)$}
    \Text(70,-50)[]{$\nu_j(k)$}
    \Text(-20,0)[]{$V_{3}^{-}$}
    \Text(28,0)[]{$p-k$}
    \Text(0,53)[]{$+i\, h_{*L}^{1jk}$}
    \Text(0,-53)[]{$+i\, h_{*L}^{1jk}$}
    \Text(0,-85)[]{$(b)$}
    \SetWidth{0.5}
    \LongArrow(12,-15)(12,15)
    \end{picture}
	\caption{
	Diagrams contributing to neutrino oscillation matter effects from the exchange of (a) $V_1^0$
	or the isospin $0$ component of $\vec{V}_3$, and (b) the isospin $-1$ component of $\vec{V}_3$. 
	The EM charges $Q_{em}=I_3+Y$ for $V_1^0$ and $V_3^0$ are $+\frac{2}{3}$, 
	while that for $V_3^{-}$ is $-\frac{1}{3}$. 
	}
	\label{LQ78}
\end{figure}

The relevant interactions for $V_1$ are
\begin{equation} \label{v11}
\mathcal{L} =
h_{1L}^{12}(\overline{u_{L}}\gamma^\mu \nu_{\mu L})V_{1\mu}
+ h_{1L}^{13}(\overline{u_{L}}\gamma^\mu \nu_{\tau L})V_{1\mu}
+ h.c.
\end{equation}
and those for $\vec V_3$  are
\begin{eqnarray} \label{v3}
\mathcal{L} & = &
h_{3L}^{12}\left[ 
(\overline{u_{L}}\gamma^\mu \nu_{\mu L})V_{3\mu}^{0}
+\sqrt{2}(\overline{d_{L}}\gamma^\mu \nu_{\mu L})V_{3\mu}^{-}\right] 
\cr & &
+ h_{3L}^{13}\left[  
(\overline{u_{L}}\gamma^\mu \nu_{\tau L})V_{3\mu}^{0}
+\sqrt{2}(\overline{d_{L}}\gamma^\mu \nu_{\tau L})V_{3\mu}^{-}\right]
+ h.c. 
\end{eqnarray}
The interactions described by Eqs.~(\ref{v11}) and (\ref{v3}) can be written in a common general form as
\begin{equation} 
\mathcal{L} = 
\lambda\,(\overline{q}\gamma^{\mu} P_L\, \nu) V +
\lambda^*(\overline{\nu}\gamma^{\mu} P_L\, q) \bar{V} \;.
\end{equation}
The Feynman diagrams contributing to neutrino oscillation matter effects are shown in Fig.~\ref{LQ78}. For momenta much smaller than the mass of the leptoquark the corresponding matrix element is
\begin{equation}
i\mathcal{M}  =  (-i)^2|\lambda|^2
\bra{\nu,q} \left( \overline{\nu}\gamma^{\mu}P_L\, q \right)
\left( \dfrac{i}{M^2_{V}} \right)
\left( \overline{q}\gamma_{\mu} P_L\, \nu \right)
\ket{\nu,q}\;.
\end{equation}
Using the Fiertz identity given in Eq.~(\ref{Fiertz2}) again, we find
\begin{equation}
i\mathcal{M}  
=  -i\frac{\displaystyle |\lambda|^2}{\displaystyle M^2_{V}}
\,\bra{\nu}\overline\nu\gamma^{\mu}P_L\,\nu \ket{\nu}
\,\bra{q}\overline q\gamma_{\mu}P_L\, q \ket{q}
\rightarrow -i\frac{|\lambda|^2}{2 M_{V}^2}N_q\left( \phi_{\nu}^{\dagger}\phi_{\nu}^{\phantom{\dagger}}\right)
=-iV_{\nu}\left( \phi_{\nu}^{\dagger}\phi_{\nu}^{\phantom{\dagger}} \right)\;,
\end{equation}
where 
\begin{equation} \label{potential-4}
V_{\nu}\equiv +\frac{N_q}{2}
\dfrac{|\lambda|^2}{M^2_{V}} \;.
\end{equation}
Applying this result to the $V_1$ case, effective potential is
\begin{equation}
V_{\nu_j} 
\;=\; +\frac{N_u}{2}\dfrac{\left|h_{1L}^{1j}\right|^2}{\left(M_{V_{1}}\right)^2}
\;=\; +\frac{(2N_p+N_n)}{2}\dfrac{\left|h_{1L}^{1j}\right|^2}{\left(M_{V_{1}}\right)^2}
\;\approx\; +\frac{3N}{2}\dfrac{\left|h_{1L}^{1j}\right|^2}{\left(M_{V_{1}}\right)^2}\;.
\end{equation}
The effective $\xi$ is
\begin{equation}
\xi_{V_1}
\;=\; \dfrac{V_{\nu_3}-V_{\nu_2}}{V_{CC}}
\;=\; -6\;
\dfrac{(\; \left| h_{1L}^{12} \right|^2 -\left| h_{1L}^{13} \right|^2 \;) / M_{V_{1}}^2 }
      {g^2/M_W^2} \;.
\end{equation}
The effective potential for the $\vec{V}_3$ case is
\begin{eqnarray}
V_{\nu_j} 
& = & +\dfrac{N_u}{2}\dfrac{\left| h_{3L}^{1j} \right|^2}{M_{V_3^0}^2}
      +N_d\,\dfrac{\left| h_{3L}^{1j} \right|^2}{M_{V_3^-}^2} \cr
& = & +\left| h_{3L}^{1j} \right|^2
\left[ \dfrac{(2N_p+N_n)}{2M_{V_3^0}^2} + \dfrac{(N_p+2N_n)}{M_{V_3^-}^2} \right] \cr
& \approx & +\dfrac{3N}{2}\,\left| h_{3L}^{1j} \right|^2
\left( \dfrac{1}{M_{V_{3}^0}^2} 
      +\dfrac{2}{M_{V_{3}^-}^2}
\right)\;.
\end{eqnarray}
The effective $\xi$ is
\begin{equation}
\xi_{\vec{V}_3}
\;=\; \dfrac{V_{\nu_3}-V_{\nu_2}}{V_{CC}}
\;=\; -6\,
\frac{\;\left| h_{3L}^{12} \right|^2 - \left| h_{3L}^{13} \right|^2\;}{g^2/M_W^2}
\left(\dfrac{1}{M_{V_{3}^0}^2}+\dfrac{2}{M_{V_{3}^-}^2}\right) \;.
\end{equation}
In the case of degenerate mass, $M_{V_3^0}=M_{V_3^-}\equiv M_{V_3}$,
we have
\begin{equation}
\xi_{\vec{V}_3}
\;=\; -18\;
\dfrac{(\; \left| h_{3L}^{12} \right|^2 - \left| h_{3L}^{13} \right|^2 \;) / M_{V_3}^2 }
          {g^2/M_W^2}\;.
\end{equation}


\begin{center}
\begin{table}[tbp] 
\begin{tabular}{|c||c|c|c|l|} 
\hline
$\;\; LQ \;\;$ 
& $\;\;\; C_{LQ} \;\;\;$ 
& $\quad\qquad\delta\lambda_{LQ}^2\qquad\quad$ 
& \ upper bound from $|\xi|\le\xi_0$\ \    
& \ current bounds from Ref.~\cite{davidson}\ \ \ \\
\hline\hline
$S_1$  
& $+3$  
& $ |g_{1L}^{12}|^2-|g_{1L}^{13}|^2$ 
& $1.1\times 10^{-3}$ 
& $\;(g^{12}_{1L})^2\le 0.008 \quad (R_{\pi})$ \\
& & &  
& $\;(g^{13}_{1L})^2\le 0.7\quad  (\tau\rightarrow\pi\nu)$ \\
\hline
$\vec S_3$       
& $+9$                  
&  $|g_{3L}^{12}|^2-|g_{3L}^{13}|^2$  
& $3.7\times 10^{-4}$ 
& $\;(g^{12}_{3L})^2\le 0.008\quad  (R_{\pi})$ \\
& & &
& $\;(g^{13}_{3L})^2\le 0.7\quad  (\tau\rightarrow\pi\nu)$ \\
\hline
$S_2$         
& $-3$                  
&  $|h_{2L}^{12}|^2-|h_{2L}^{13}|^2$  
& $1.1\times 10^{-3}$ 
& $\;(h^{12}_{2L})^2\le 1 \quad (\mu N\rightarrow\mu X)$ \\
\hline
$\tilde{S}_2$   
& $-3$                  
& $|\tilde{h}{}_{2L}^{12}|^2-|\tilde{h}{}_{2L}^{13}|^2$  
& $1.1\times 10^{-3}$ 
& $\;(\tilde h^{12}_{2L})^2\le 2 \quad (\mu N\rightarrow\mu X)$ \\
\hline
$V_2$          
& $+6$                  
& $|g_{2L}^{12}|^2-|g_{2L}^{13}|^2$  
& $5.5\times 10^{-4}$ 
& $\;(g^{12}_{2L})^2\le 1 \quad (\mu N\rightarrow\mu X)$ \\
\hline
$\tilde{V}_2$   
& $+6$                  
& $|\tilde{g}{}_{2L}^{12}|^2-|\tilde{g}{}_{2L}^{13}|^2$  
& $5.5\times 10^{-4}$ 
& $\;(\tilde g^{12}_{2L})^2\le 5 \quad (\mu N\rightarrow\mu X)$ \\
\hline
$V_1$            
& $-6$                  
& $|h_{1L}^{12}|^2-|h_{1L}^{13}|^2$ 
& $5.5\times 10^{-4}$ 
& $\;(h^{12}_{1L})^2\le 0.004 \quad (R_{\pi})$ \\
& & &
& $\;(h^{13}_{1L})^2\le 0.1 \quad (D\rightarrow\mu\nu)$ \\
\hline
$\vec V_3$       
& $-18$                 
& $|h_{3L}^{12}|^2-|h_{3L}^{13}|^2$  
& $1.8\times 10^{-4}$ 
& $\;(h^{12}_{3L})^2\le 0.004 \quad (R_{\pi})$ \\
& & &
& $\;(h^{13}_{3L})^2\le 0.1\quad  (D\rightarrow\mu\nu)$ \\
\hline
\end{tabular}
\caption[]{Constraints on the leptoquark couplings with all the leptoquark masses set to
100~GeV. To obtain the bounds for a different leptoquark mass $M_{LQ}$, simply
rescale these numbers with the factor $(M_{LQ}/100\text{ GeV})^2$.}
\label{tab3}
\end{table}
\end{center}

\begin{figure}[ht]
\centering
\includegraphics[width=8.1cm]{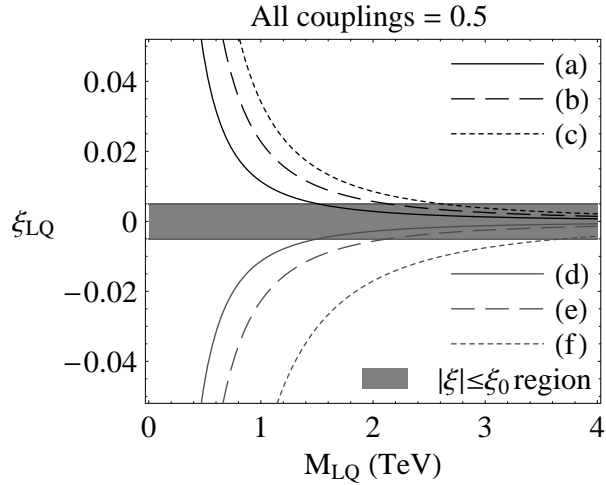}		
\caption{
$\xi_{LQ}$ dependence on the leptoquark mass for 
$\sqrt{\Delta\lambda^2_{LQ}}=0.5$. 
(a) $S_1$; 
(b) $V_2$, $\tilde{V}_2$;
(c) $\vec{S}_3$; 
(d) $S_2$, $\tilde{S}_2$; 
(e) $V_1$; 
(f) $\vec{V}_3$. 
}
\label{LQ_result_xi_mass}
\end{figure}
\begin{figure}[ht]
\centering
\includegraphics[width=8.1cm]{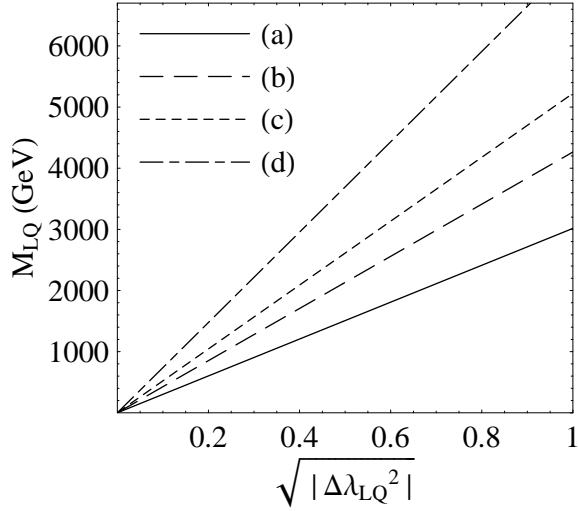}		
\caption{Lower bounds on the leptorquark masses. 
(a) $S_1$, $S_2$, $\tilde{S}_2$; 
(b) $V_1$, $V_2$, $\tilde{V}_2$; 
(c) $\vec{S}_3$; (d) $\vec{V}_3$. }
\label{LQ_result}
\end{figure}

\begin{center}
\begin{table}[t]
\begin{tabular}{|l||c|c|c|c|c|}
\hline
Process 
& \ $(ij)$\ \ 
& \ LQ\ \ 
& \ Assumptions\ \ 
& \ 95\% CL bound\ \ 
& \ Reference\ \ \\
\hline
$\;p\bar{p}\rightarrow LQ\,\overline{LQ}\,X \rightarrow (j\nu)(j\nu)X\;$ 
& $(**)$
& $S$
& $\beta=0^{(a)}$
& $117\;\mathrm{GeV}^{\phantom{(a)}}$
& CDF \cite{Acosta:2004zb}
\\
\hline
$\;p\bar{p}\rightarrow LQ\,\overline{LQ}\,X \rightarrow (j\nu)(j\nu)X\;$ 
& $(**)$
& $S$
& $\beta=0^{\phantom{(a)}}$ 
& $135\;\mathrm{GeV}^{\phantom{(a)}}$
& D0 \cite{Abazov:2006wp}
\\
\hline
$\;p\bar{p}\rightarrow LQ\,\overline{LQ}\,X \rightarrow (j\mu)(j\mu)X\;$ 
& $\;(*2)\;$
& $S$
& $\beta=0.5$
& $208\;\mathrm{GeV}^{\phantom{(a)}}$
& CDF \cite{Abulencia:2005ua}
\\
$\;p\bar{p}\rightarrow LQ\,\overline{LQ}\,X \rightarrow (j\mu)(j\nu)X\;$ 
& 
& 
&
& 
& 
\\
\hline
$\;p\bar{p}\rightarrow LQ\,\overline{LQ}\,X \rightarrow (j\mu)(j\mu)X\;$ 
& $\;(*2)\;$
& $S$
& $\beta=0.5$
& $204\;\mathrm{GeV}^{\phantom{(a)}}$
& D0 \cite{Abazov:2006vc}
\\
$\;p\bar{p}\rightarrow LQ\,\overline{LQ}\,X \rightarrow (j\mu)(j\nu)X\;$ 
& 
& 
&
& 
& 
\\
\hline
$\;p\bar{p}\rightarrow LQ\,\mu\,X \rightarrow (j\mu)\mu\,X\;$
& $\;(*2)\;$
& $S$
& $\;\beta=0.5,\;\lambda=1^{(b)}\;$
& $226\;\mathrm{GeV}^{(c)}$
& D0 \cite{Abazov:2006ej}
\\
\hline
$\;p\bar{p}\rightarrow LQ\,\overline{LQ}\,X \rightarrow (j\tau)(j\tau)X\;$ 
& $\;(*3)\;$
& $V$
& \ minimal coupling \cite{Blumlein:1996qp}\ \
& $251\;\mathrm{GeV}^{\phantom{(a)}}$
& CDF \cite{CDF8309}
\\
\hline
\end{tabular}
\caption{Direct search limits on the Leptoquark mass from the Tevatron.  
${}^{(a)}$$\beta$ is the assumed branching fraction $B(LQ\rightarrow q\ell) = 1 - B(LQ\rightarrow q\nu)$, and 
${}^{(b)}$$\lambda$ is the Yukawa coupling of the Leptoquark with the quark-lepton pair.  ${}^{(c)}$Combined bound with the pair production data. 
}
\label{TevatronLQmassBounds}
\end{table}
\end{center}

\subsection{Constraints on the Leptoquark Couplings and Masses}

Assuming a common mass for leptoquarks in the same $SU(2)_L$ weak-isospin multiplet,
the effective $\xi$ due to the exchange of any particular type of leptoquark can be written in the form
\begin{equation} \label{xi_leptoquark}
\xi_{LQ}=C_{LQ}\;\frac{\;\delta\lambda_{LQ}^2/M_{LQ}^2\;}{g^2/M_W^2}=
\frac{C_{LQ}}{4\sqrt{2} G_F}\left(\frac{\delta\lambda^2_{LQ}}{M^2_{LQ}}\right)
\;.
\end{equation}
Here, $C_{LQ}$ is a constant prefactor, and $\delta\lambda_{LQ}^2$ represents
\begin{equation}
\delta\lambda_{LQ}^2 = |\lambda_{LQ}^{12}|^2 - |\lambda_{LQ}^{13}|^2\;,
\end{equation}
where $\lambda_{LQ}^{ij}$ is a generic coupling constant.
The values of $C_{LQ}$ and $\delta\lambda_{LQ}^2$ for the different types of leptoquark are 
listed in the first two columns of Table~\ref{tab3}.
In Fig.~\ref{LQ_result_xi_mass}, we show how $\xi_{LQ}$ 
depend on the leptoquark mass $M_{LQ}$ for the choice 
$\sqrt{\delta\lambda^2_{LQ}}=0.5$,
where we have assumed $\delta\lambda^2_{LQ} > 0$.
To obtain the picture for the case when 
$\delta\lambda^2_{LQ} < 0$, the vertical axis of the graph 
should be flipped.
The constraint $|\xi_{LQ}|\le\xi_0$ translates into:
\begin{equation}
M_{LQ} \ge M_W
\sqrt{\dfrac{|\delta\lambda_{LQ}^2|}{g^2}}\sqrt{\frac{|C_{LQ}|}{\xi_0}}
= \sqrt{\dfrac{|C_{LQ}| |\delta\lambda_{LQ}^2|}{4\sqrt{2}G_F\,\xi_0}}
\approx \sqrt{|C_{LQ}| |\delta\lambda_{LQ}^2|}\times (1700\,\mathrm{GeV}) \;.
\label{LQmassBound}
\end{equation}
The resulting bounds are shown in Fig.~\ref{LQ_result}, where 
the regions of the $(M_{LQ}, \sqrt{|\delta\lambda_{LQ}^2|})$ parameter space
below each of the lines will be excluded.
One can also fix the leptoquark mass and obtain upper bounds on the leptoquark couplings:
\begin{equation}
|\delta\lambda^2_{LQ}| 
\;\le\; \left(\dfrac{4\sqrt{2}G_F\,\xi_0}{|C_{LQ}|}\right) M^2_{LQ}
\;=\; \dfrac{\;3.3\times 10^{-3}\;}{|C_{LQ}|}\left(\dfrac{M_{LQ}}{100\,\mathrm{GeV}}\right)^2\;.
\end{equation}
The values when $M_{LQ}=100\,\mathrm{GeV}$ are listed in the third column of
Table~\ref{tab3}.  
The bounds for a different choice of leptoquark mass $M_{LQ}$ 
can be obtained by multiplying by a factor of $(M_{LQ}/100\mathrm{GeV})^2$. 
This result can be compared with various indirect bounds from 
rare processes which are listed in the last column of Table~\ref{tab3}.
As can be seen, 
the limits from $|\xi|\le\xi_0$ can significantly improve existing bounds.

Limits on leptoquark masses from direct searches at the Tevatron
are listed in Table~\ref{TevatronLQmassBounds}.  Bounds from LEP and LEP~II
are weaker due to their smaller center of mass energies.
Since neutrino oscillation is only sensitive to leptoquarks with $(ij)=(12)$ and/or $(ij)=(13)$ couplings, we only quote limits which apply to
leptoquarks with \textit{only} those particular couplings, that is, 
leptoquarks that decay into a first generation quark, 
and either a second or third generation lepton.
Though it is usually stated in collider analyses 
that leptoquarks are assumed to decay into a quark-lepton pair of one particular 
generation, it is often the case that the jets coming from the quarks are not
flavor tagged.  Analyses that look for the leptoquark in the quark-neutrino decay channel are of course blind to the flavor of the neutrino.
Therefore, the bounds listed apply to leptoquarks with generation non-diagonal couplings also.

As can be seen from Table~\ref{TevatronLQmassBounds}, 
the mass bounds from the Tevatron are typically around $200\,\mathrm{GeV}$ and
are mostly independent of the leptoquark-quark-lepton coupling $\lambda$.
This independence is due to the dominance of 
the strong interaction processes, $q\bar{q}$ annihilation and gluon fusion,
in the leptoquark pair-production cross sections, and the fact that
heavy leptoquarks decay without a displaced vertex even for very small
values of $\lambda$:
the decay widths of scalar and vector leptoquarks with leptoquark-quark-lepton
coupling $\lambda$ are given by $\lambda^2 M_{LQ}/16\pi$ and
$\lambda^2 M_{LQ}/24\pi$, respectively, which correspond to lifetimes of
$O(10^{-21})$ seconds for $M_{LQ}=O(10^2)$~GeV, and $\lambda=O(10^{-2})$. 
In contrast, the potential bound on $M_{LQ}$ from neutrino oscillation,
Eq.~(\ref{LQmassBound}), depends on the coupling 
$\sqrt{|C_{LQ}||\delta\lambda^2_{LQ}|}$, 
but can be expected to be stronger than the existing ones for 
$\sqrt{|C_{LQ}||\delta\lambda^2_{LQ}|}$ as small as $0.1$.

Bounds on leptoquarks with $(ij)=(12)$ couplings can also be obtained from 
bounds on contact interactions of the form
\begin{equation}
\mathcal{L} = \pm
\dfrac{4\pi}{(\Lambda_{q\mu}^\pm)^2}
\left(\bar{q}\gamma^\mu P_{X} q\right)
\left(\bar{\mu}\gamma_\mu P_L\mu\right)
\;,
\end{equation}
where $X=L$ or $R$, and $q=u$ or $d$.
For instance, at energies much lower than the leptoquark mass,
the exchange of the $S_1$ leptoquark leads to the interaction
\begin{equation}
\mathcal{L}_{S_1} = +\dfrac{|g_{1L}^{12}|^2}{2M_{S_1}^2}
\left(\bar{u}\gamma^\mu P_L u\right)\left(\bar{\mu}\gamma_\mu P_L \mu\right)\;.
\end{equation}
The remaining cases are listed in Table~\ref{LQcontact}.
The 95\% CL lower bounds on the $\Lambda^\pm_{q\ell}$'s from CDF can be found in 
Ref.~\cite{Abe:1997gt}, and the cases relevant to our discussion 
are listed in Table~\ref{TevatronContact}.
These bounds translate into bounds on the leptoquark masses and couplings 
listed in Table~\ref{LQcontact}. 
Clearly, the potential bounds from $|\xi|<\xi_0$, also listed in
Table~\ref{LQcontact}, are much stronger.
It should be noted, though, that the results of Ref.~\cite{Abe:1997gt} are 
from Tevatron Run I, and we can expect the Run II results to improve these 
bounds.  Indeed, Ref.~\cite{Xuan:2005gw} from D0, which we cited earlier in the
$Z'$ section, analyzes the Run II data for contact interactions of the
form
\begin{equation}
\mathcal{L} = \pm
\dfrac{4\pi}{(\Lambda^\pm)^2}
\left(\bar{u}\gamma^\mu P_{X} u + \bar{d}\gamma^\mu P_{X} d\right)
\left(\bar{\mu}\gamma_\mu P_L\mu\right)
\;,\qquad \mbox{$X=L$ or $R$}\;,
\end{equation}
and places 95\% CL lower bounds on the 
$\Lambda^\pm$'s in the $4\sim 7$ TeV range.
While these are not exactly the interactions induced by leptoquarks, 
we can nevertheless expect that the bounds on the $\Lambda^\pm_{q\mu}$'s
will be in a similar range, and thereby conclude that the 
Run II data will roughly double the lower bounds from Run I.
Even then, Table~\ref{LQcontact} indicates that the potential bounds
from $|\xi|<\xi_0$ will be much stronger.

\begin{table}[t]
\begin{tabular}{|c|c|c|c|}
\hline
\ LQ\ \ 
& Induced Interaction 
& \ CDF 95\% CL \cite{Abe:1997gt}\ \ 
& \ $|\xi|<\xi_0$\ \ \\
\hline\hline
$S_1$ & 
$+\dfrac{|g_{1L}^{12}|^2}{2M_{S_1}^2}
\left(\bar{u}\gamma^\mu P_L u\right)
\left(\bar{\mu}\gamma_\mu P_L \mu\right)$ 
& \ $\dfrac{M_{S_1}}{|g_{1L}^{12}|}\ge 0.68\,\mathrm{TeV}$\ \ 
& \ $\dfrac{M_{S_1}}{\sqrt{\delta g_{1L}^2}}\ge 3.0\,\mathrm{TeV}$\ \
\\ 
\hline
$S_2$ &
$-\dfrac{|h_{2L}^{12}|^2}{2M_{S_2}^2}
\left(\bar{u}\gamma^\mu P_R u\right)
\left(\bar{\mu}\gamma_\mu P_L \mu\right)$ 
& \ $\dfrac{M_{S_2}}{|h_{2L}^{12}|}\ge 0.72\,\mathrm{TeV}$\ \
& \ $\dfrac{M_{S_2}}{\sqrt{\delta h_{2L}^2}}\ge 3.0\,\mathrm{TeV}$\ \
\\
\hline
$\tilde{S}_2$ &
$-\dfrac{|\tilde{h}_{2L}^{12}|^2}{2M_{\tilde{S}_2}^2}
\left(\bar{d}\gamma^\mu P_R d\right)
\left(\bar{\mu}\gamma_\mu P_L \mu\right)$ 
& \ $\dfrac{M_{\tilde{S}_2}}{|\tilde{h}_{2L}^{12}|}\ge 0.38\,\mathrm{TeV}$\ \
& \ $\dfrac{M_{\tilde{S}_2}}{\sqrt{\delta \tilde{h}_{2L}^2}}\ge 3.0\,\mathrm{TeV}$\ \
\\
\hline
$S_3$ & 
\ $+\dfrac{|g_{3L}^{12}|^2}{2M_{S_3}^2}
\left(\bar{u}\gamma^\mu P_L u + 2\,\bar{d}\gamma^\mu P_L d\right)
\left(\bar{\mu}\gamma_\mu P_L \mu\right)$\ \ 
& ---
& \ $\dfrac{M_{\tilde{S}_3}}{\sqrt{\delta \tilde{g}_{3L}^2}}\ge 5.2\,\mathrm{TeV}$\ \
\\
\hline
$V_1$ &
$-\dfrac{|h_{1L}^{12}|^2}{M_{V_1}^2}
\left(\bar{d}\gamma^\mu P_L d\right)
\left(\bar{\mu}\gamma_\mu P_L \mu\right)$
& \ $\dfrac{M_{V_1}}{|h_{1L}^{12}|}\ge 0.48\,\mathrm{TeV}$\ \
& \ $\dfrac{M_{V_1}}{\sqrt{\delta h_{1L}^2}}\ge 4.3\,\mathrm{TeV}$\ \
\\
\hline
$V_2$ &
$+\dfrac{|g_{2L}^{12}|^2}{M_{V_2}^2}
\left(\bar{d}\gamma^\mu P_R d\right)
\left(\bar{\mu}\gamma_\mu P_L \mu\right)$ 
& \ $\dfrac{M_{V_2}}{|g_{2L}^{12}|}\ge 0.56\,\mathrm{TeV}$\ \
& \ $\dfrac{M_{V_2}}{\sqrt{\delta g_{2L}^2}}\ge 4.3\,\mathrm{TeV}$\ \
\\
\hline
$\tilde{V}_2$ &
$+\dfrac{|\tilde{g}_{2L}^{12}|^2}{M_{\tilde{V}_2}^2}
\left(\bar{u}\gamma^\mu P_R u\right)
\left(\bar{\mu}\gamma_\mu P_L \mu\right)$ 
& \ $\dfrac{M_{\tilde{V}_2}}{|\tilde{g}_{2L}^{12}|}\ge 0.85\,\mathrm{TeV}$\ \
& \ $\dfrac{M_{\tilde{V}_2}}{\sqrt{\delta \tilde{g}_{2L}^2}}\ge 4.3\,\mathrm{TeV}$\ \
\\
\hline
$V_3$ &
$-\dfrac{|h_{3L}^{12}|^2}{M_{V_1}^2}
\left(2\,\bar{u}\gamma^\mu P_L u +\bar{d}\gamma^\mu P_L d\right)
\left(\bar{\mu}\gamma_\mu P_L \mu\right)$ 
& ---
& \ $\dfrac{M_{\tilde{V}_3}}{\sqrt{\delta \tilde{h}_{3L}^2}}\ge 7.4\,\mathrm{TeV}$\ \
\\
\hline
\end{tabular}
\caption{The quark-muon interactions induced by leptoquark exchange, and the
bounds from CDF \cite{Abe:1997gt} compared with potential bounds from
neutrino oscillations.
Only the couplings that also contribute to neutrino oscillation are listed.
Analysis of the Tevatron Run II data is expected to improve the
CDF bound by a factor of two.}
\label{LQcontact}
\end{table}

\begin{table}[t]
\begin{tabular}{|c||c|c|c|c|}
\hline
\ $(q\mu)$ chirality\ \
& \ $\Lambda_{u\mu}^+$ (TeV)\ \ 
& \ $\Lambda_{u\mu}^-$ (TeV)\ \ 
& \ $\Lambda_{d\mu}^+$ (TeV)\ \
& \ $\Lambda_{d\mu}^-$ (TeV)\\
\hline
\ $(LL)$\ \ & 3.4 & 4.1 & 2.3 & 1.7 \\
\ $(RL)$\ \ & 3.0 & 3.6 & 2.0 & 1.9 \\ 
\hline
\end{tabular}
\caption{The 95\% CL lower bound on the compositeness scale
from CDF \cite{Abe:1997gt}.  Results from D0 \cite{Xuan:2005gw}
do not provide limits for cases where the muons couple to
only $u$ or $d$, but we expect the bounds to be in the range $4\sim 7$~TeV.}
\label{TevatronContact}
\end{table}

The prospects for leptoquark discovery at the LHC are discussed in
Refs.~\cite{LHC-TDR,Mitsou:2004hm}.
At the LHC, leptoquarks can be pair-produced via gluon fusion and 
quark-antiquark annihilation, or singly-produced with an accompanying 
lepton via quark-gluon fusion.
The pair-production cross section is dominated by gluon fusion, which does not involve the leptoquark-quark-lepton coupling $\lambda$, 
and is therefore independent of the details assumed for the leptoquark interactions.
Once produced, each leptoquark will decay into a lepton plus
jet, regardless of
whether the coupling is generation diagonal or not.
The leptoquark width in this decay depends on $\lambda$, but it is
too narrow compared to the calorimeter resolution 
for the $\lambda$-dependence to be of relevance in the analyses. 
Therefore, though the analyses of Refs.~\cite{LHC-TDR,Mitsou:2004hm}
assume specific values of $\lambda$ and generation diagonal couplings, we expect their conclusions to apply equally well to different $\lambda$-values
and generation non-diagonal cases:
for $\beta=B(LQ\rightarrow q\ell)=0.5$, the expected sensitivity is up to
$M_{LQ} \approx 1\,\mathrm{TeV}$ with 30${}^{-1}\;\mathrm{fb}$ of data 
\cite{Mitsou:2004hm}.
Again, in contrast, the 
the potential bound from neutrino oscillation,
Eq.~(\ref{LQmassBound}), depends on the coupling 
$\sqrt{|C_{LQ}||\delta\lambda^2_{LQ}|}$.
If $\sqrt{|C_{LQ}||\delta\lambda^2_{LQ}|} = O(1)$, then 
Eq.~(\ref{LQmassBound}) will be competitive with the expected LHC bound.


\section{SUSY Standard Model with R-parity Violation}


Let us next consider contributions from R-parity violating couplings.
Assuming the particle content of the 
Minimal Supersymmetric Standard Model (MSSM),
the most general R-parity violating superpotential (involving only
tri-linear couplings) has the form
\cite{Rparity_notations}
\begin{equation}
W_{\not R}
=\frac{1}{2}\lambda_{ijk}\hat{L}_i \hat{L}_j \hat{E}_k
+\lambda^{\prime}_{ijk}\hat{L}_i \hat{Q}_j \hat{D}_k
+\frac{1}{2}\lambda^{\prime\prime}_{ijk}\hat{U}_i \hat{D}_j \hat{D}_k\;,
\label{RPVlagrangian}
\end{equation}
where $\hat{L}_i$, $\hat{E}_i$, $\hat{Q}_i$, $\hat{D}_i$, and $\hat{U}_i$ are
the left-handed MSSM superfields defined in the usual fashion,
and the subscripts $i,j,k=1,2,3$ are the generation indices.
(Note, however, that in some references, such as Ref.~\cite{Barbier:2004ez}, 
the isospin singlet superfields $\hat{E}_i$, $\hat{D}_i$, and
$\hat{U}_i$ are defined to be right-handed, so the corresponding 
left-handed fields in Eq.~(\ref{RPVlagrangian})
appear with a superscript $c$ indicating charge-conjugation.)
$SU(2)_L$ gauge invariance requires
the couplings $\lambda_{ijk}$ to be antisymmetric in the first two indices:
\begin{equation}
\lambda_{ijk} \;=\; -\lambda_{jik}\;,
\end{equation}
whereas $SU(3)$ gauge invariance requires 
the couplings $\lambda''_{ijk}$ to be antisymmetric in the latter two:
\begin{equation}
\lambda''_{ijk} \;=\; -\lambda''_{ikj}\;.
\end{equation}
These conditions reduce the number of R-parity violating couplings in 
Eq.~(\ref{RPVlagrangian}) to 45 (9 $\lambda_{ijk}$, 27 $\lambda'_{ijk}$,
and 9 $\lambda''_{ijk}$).
The purely baryonic operator $\hat{U}_i\hat{D}_j\hat{D}_k$ is irrelevant to our discussion on neutrino oscillation so we will not consider the
$\lambda''_{ijk}$ couplings further.
We also neglect possible bilinear R-parity violating couplings which have the
effect of mixing the neutrinos with the neutral higgsino; their effect
on neutrino oscillation has been discussed extensively by many authors \cite{Barbier:2004ez,Aulakh:1982yn,Hempfling:1995wj}.


\subsection{$\hat{L}\hat{L}\hat{E}$ couplings}


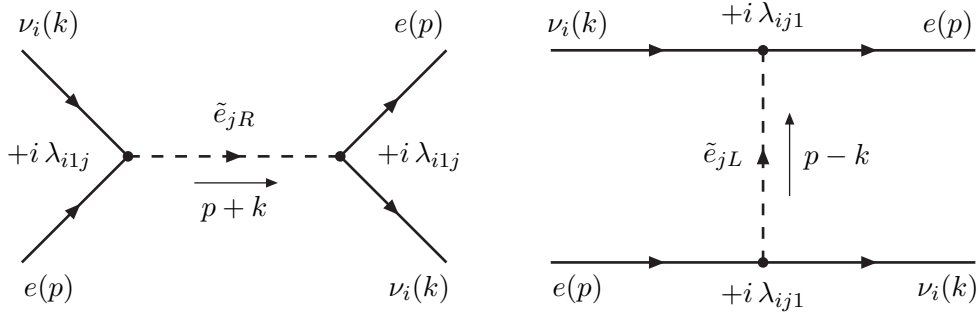
\begin{figure}[ht]
\centering
    \begin{picture}(400,120)(-100,-60) 
    \SetScale{1}  
    \SetColor{Black}
    \SetWidth{1}
    \ArrowLine(-80,-40)(-40,0)
    \ArrowLine(-80,40)(-40,0)
    \DashArrowLine(-40,0)(40,0){5}
    \ArrowLine(40,0)(80,40)
    \ArrowLine(40,0)(80,-40)
    \Vertex(-40,0){2}
    \Vertex(40,0){2}       
    \Text(-70,50)[]{$\nu_i(k)$}
    \Text(70,50)[]{$e(p)$}
    \Text(-70,-50)[]{$e(p)$}
    \Text(70,-50)[]{$\nu_i(k)$}
    \Text(0,15)[]{$\tilde{e}_{jR}$}
    \Text(0,-20)[]{$p+k$}
    \Text(-70,0)[]{$+i\,\lambda_{i1j}$}
    \Text(70,0)[]{$+i\,\lambda_{i1j}$}    
    \SetWidth{0.5}
    \LongArrow(-15,-10)(15,-10)
    
    \SetOffset(200,0)
    \SetWidth{1}
    \ArrowLine(-80,-40)(0,-40)
    \ArrowLine(0,-40)(80,-40)
    \DashArrowLine(0,-40)(0,40){5}
    \ArrowLine(-80,40)(0,40)
    \ArrowLine(0,40)(80,40)
    \Vertex(0,40){2}
    \Vertex(0,-40){2}       
    \Text(-70,50)[]{$\nu_i(k)$}
    \Text(70,50)[]{$e(p)$}
    \Text(-70,-50)[]{$e(p)$}
    \Text(70,-50)[]{$\nu_i(k)$}
    \Text(-15,0)[]{$\tilde{e}_{jL}$}
    \Text(28,0)[]{$p-k$}
    \Text(0,53)[]{$+i\, \lambda_{ij1}$}
    \Text(0,-53)[]{$+i\, \lambda_{ij1}$}
    \SetWidth{0.5}
    \LongArrow(10,-15)(10,15)

    \end{picture}
	\caption{$LLE$ interactions that contribute to neutrino oscillation matter effects..
	}
	\label{Rparity-LLE}
\end{figure}

The $\hat{L}\hat{L}\hat{E}$ part of the  R-parity violating Lagrangian,
Eq.~(\ref{RPVlagrangian}), expressed in terms of the component fields is 
\begin{equation}
\mathcal{L}_{LLE} \;=\; \lambda_{ijk}\left[
\tilde{\nu}_{iL}\overline{e_{kR}}e_{jL}
+\tilde{e}_{jL}\overline{e_{kR}} \nu_{iL}
+\tilde{e}_{kR}^{*}\overline{\nu_{iL}^c} e_{jL}
\right]+h.c. 
\end{equation}
The second and third terms of this Lagrangian, together with their hermitian conjugates, contribute to neutrino oscillation matter effects. 
The corresponding Feynman diagrams are shown in Fig~\ref{Rparity-LLE}.
Since $\lambda_{ijk}$ is antisymmetric under 
$i \leftrightarrow j$, it follows that $i \ne j$. 
Calculations similar to those for the scalar leptoquarks yield 
\begin{equation}
V_{\tilde{e}}(\nu_i)
=\dfrac{N_e}{4}\left( 
 \sum_{j\neq i} \dfrac{|\lambda_{ij1}|^2}{M_{\tilde{e}_{jL}}^2}
-\sum_{j} \dfrac{|\lambda_{i1j}|^2}{M_{\tilde{e}_{jR}}^2}
\right) \;,
\end{equation}
or if we write everything out explicitly:
\begin{eqnarray}
V_{\tilde{e}}(\nu_2)
& = & \dfrac{N_e}{4}\left(
\dfrac{|\lambda_{211}|^2}{M_{\tilde{e}_{1L}}^2}
+\dfrac{|\lambda_{231}|^2}{M_{\tilde{e}_{3L}}^2}
-\dfrac{|\lambda_{211}|^2}{M_{\tilde{e}_{1R}}^2}
-\dfrac{|\lambda_{212}|^2}{M_{\tilde{e}_{2R}}^2}
-\dfrac{|\lambda_{213}|^2}{M_{\tilde{e}_{3R}}^2} 
\right)\;,\cr
V_{\tilde{e}}(\nu_3)
& = & \dfrac{N_e}{4}\left(
\dfrac{|\lambda_{311}|^2}{M_{\tilde{e}_{1L}}^2}
+\dfrac{|\lambda_{321}|^2}{M_{\tilde{e}_{2L}}^2}
-\dfrac{|\lambda_{311}|^2}{M_{\tilde{e}_{1R}}^2}
-\dfrac{|\lambda_{312}|^2}{M_{\tilde{e}_{2R}}^2}
-\dfrac{|\lambda_{313}|^2}{M_{\tilde{e}_{3R}}^2} 
\right)\;.
\end{eqnarray}
The effective $\xi$ is
\begin{eqnarray} 
\xi_{\tilde{e}}
& = & \dfrac{V_{\tilde{e}}(\nu_3)-V_{\tilde{e}}(\nu_2)}{V_{CC}} \cr
& = & \dfrac{1}{g^2/M_W^2}
\left(
-\sum_{j=1,3}\dfrac{\left|\lambda_{2j1}\right|^2}{M^2_{\tilde{e}_{jL}}}
-\sum_{j=1,2}\dfrac{\left|\lambda_{3j1}\right|^2}{M^2_{\tilde{e}_{jL}}}
+\sum_{j=1}^3\dfrac{\left|\lambda_{21j}\right|^2-\left|\lambda_{31j}\right|^2}
      {M^2_{\tilde{e}_{jR}}}
\right) \cr
& = & \dfrac{1}{g^2/M_W^2}\left[
\left(\;\left|\lambda_{211}\right|^2-\left|\lambda_{311}\right|^2\;\right)
\left( \dfrac{1}{M^2_{\tilde{e}_{1R}}} -\dfrac{1}{M^2_{\tilde{e}_{1L}}} \right)
\right. \cr
& & \left.\qquad
+\left|\lambda_{231}\right|^2 
\left( \dfrac{1}{M^2_{\tilde{e}_{2L}}} - \dfrac{1}{M^2_{\tilde{e}_{3L}}} \right)
+\dfrac{\left|\lambda_{212}\right|^2-\left|\lambda_{312}\right|^2}{M^2_{\tilde{e}_{2R}}}
+\dfrac{\left|\lambda_{213}\right|^2-\left|\lambda_{313}\right|^2}{M^2_{\tilde{e}_{3R}}}
\right]\;.
\label{xi_LLE}
\end{eqnarray}
For degenerate s-electron masses $M_{\tilde{e}_{jL}}=M_{\tilde{e}_{jR}}\equiv M_{\tilde{e}_{j}}$, we have
\begin{equation}
\xi_{\tilde{e}}
= \dfrac{1}{g^2/M_W^2}
\left( \dfrac{
\left|\lambda_{231}\right|^2
+\left|\lambda_{122}\right|^2
-\left|\lambda_{132}\right|^2}{M^2_{\tilde{e}_{2}}}
- \dfrac{
\left|\lambda_{231}\right|^2
-\left|\lambda_{123}\right|^2
+\left|\lambda_{133}\right|^2}{M^2_{\tilde{e}_{3}}}
\right)\;,
\label{xi_LLE_degenerate}
\end{equation}
where we have used $\lambda_{ijk}=-\lambda_{jik}$ to reorder the
indices.


\subsection{$\hat{L}\hat{Q}\hat{D}$ couplings}


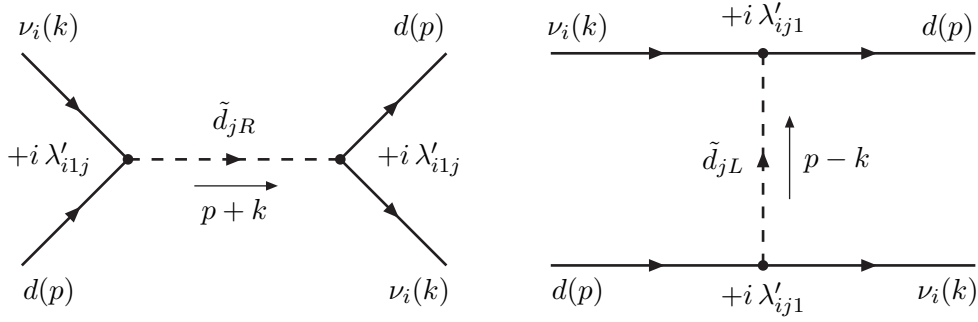
\begin{figure}[ht]
\centering
    \begin{picture}(400,120)(-100,-60) 
    \SetScale{1}  
    \SetColor{Black}
    
    \SetWidth{1}
    \ArrowLine(-80,-40)(-40,0)
    \ArrowLine(-80,40)(-40,0)
    \DashArrowLine(-40,0)(40,0){5}
    \ArrowLine(40,0)(80,40)
    \ArrowLine(40,0)(80,-40)
    \Vertex(-40,0){2}
    \Vertex(40,0){2}       
    \Text(-70,50)[]{$\nu_i(k)$}
    \Text(70,50)[]{$d(p)$}
    \Text(-70,-50)[]{$d(p)$}
    \Text(70,-50)[]{$\nu_i(k)$}
    \Text(0,15)[]{$\tilde{d}_{jR}$}
    \Text(0,-20)[]{$p+k$}
    \Text(-70,0)[]{$+i\,\lambda'_{i1j}$}
    \Text(70,0)[]{$+i\,\lambda'_{i1j}$}    
    \SetWidth{0.5}
    \LongArrow(-15,-10)(15,-10)
    
    \SetOffset(200,0)
    \SetWidth{1}
    \ArrowLine(-80,-40)(0,-40)
    \ArrowLine(0,-40)(80,-40)
    \DashArrowLine(0,-40)(0,40){5}
    \ArrowLine(-80,40)(0,40)
    \ArrowLine(0,40)(80,40)
    \Vertex(0,40){2}
    \Vertex(0,-40){2}       
    \Text(-70,50)[]{$\nu_i(k)$}
    \Text(70,50)[]{$d(p)$}
    \Text(-70,-50)[]{$d(p)$}
    \Text(70,-50)[]{$\nu_i(k)$}
    \Text(-15,0)[]{$\tilde{d}_{jL}$}
    \Text(28,0)[]{$p-k$}
    \Text(0,53)[]{$+i\, \lambda'_{ij1}$}
    \Text(0,-53)[]{$+i\, \lambda'_{ij1}$}
    \SetWidth{0.5}
    \LongArrow(10,-15)(10,15)
    \end{picture}
	\caption{$LQD$ interactions that contribute to neutrino oscillation matter effects..
	}
	\label{Rparity-LQD}
\end{figure}    

The $\hat{L}\hat{Q}\hat{D}$ part of the R-parity violating Lagrangian expressed in terms of the component fields is 
\begin{eqnarray}
\mathcal{L}_{LQD} & = & \lambda_{ijk}^{\prime}\left[ 
\tilde{\nu}_{iL}\overline{d_{kR}} d_{jL}
+\tilde{d}_{jL}\overline{d_{kR}} \nu_{iL}
+\tilde{d}_{kR}^{*}\overline{\nu_{iL}^c} d_{jL}\right. \cr
& - & \left. \left(
\tilde{e}_{iL}\overline{d_{kR}}u_{jL}
+\tilde{u}_{jL} \overline{d_{kR}} e_{iL}
+\tilde{d}_{kR}^* \overline{e^c_{iL}} u_{jL}
\right)\right]+ h.c. 
\end{eqnarray}
The second and third terms of this Lagrangian, together with their hermitian 
conjugates, contribute to neutrino oscillation matter effects. 
The corresponding Feynman diagrams are shown in Fig~\ref{Rparity-LQD}. 
Calculations similar to those for the scalar leptoquarks lead to the 
following effective potential for neutrino flavor $\nu_i$:
\begin{equation}
V_{\tilde{d}}(\nu_i)
\;=\; \sum_{j=1}^3\dfrac{N_p+2N_n}{4}
\left(
 \dfrac{\left|\lambda^\prime_{ij1}\right|^2}{M_{\tilde{d}_{jL}^2}}
-\dfrac{\left|\lambda^\prime_{i1j}\right|^2}{M_{\tilde{d}_{jR}^2}}
\right) 
\;\approx\;
\sum_{j=1}^3\dfrac{3N}{4}
\left(
 \dfrac{\left|\lambda^\prime_{ij1}\right|^2}{M_{\tilde{d}_{jL}^2}}
-\dfrac{\left|\lambda^\prime_{i1j}\right|^2}{M_{\tilde{d}_{jR}^2}}
\right) 
\;.
\end{equation}
The effective $\xi$ is
\begin{eqnarray} 
\xi_{\tilde{d}}
& = & \dfrac{V_{\tilde{d}}(\nu_3)-V_{\tilde{d}}(\nu_2)}{V_{CC}} \cr
& = & -3\sum_{j=1}^3
\dfrac{
\left(
\left|\lambda^\prime_{2j1}\right|^2
-\left|\lambda^\prime_{3j1}\right|^2
\right)/M_{\tilde d_{jL}}^2
-\left(
\left|\lambda^{\prime}_{21j}\right|^2
-\left|\lambda^\prime_{31j}\right|^2
\right)/M_{\tilde d_{jR}}^2}
{g^2/M_W^2} \;.
\label{xi_LQD}
\end{eqnarray}
For degenerate $d$-squark masses 
$M_{\tilde{d}_{jL}}=M_{\tilde{d}_{jR}}\equiv M_{\tilde{d}_j}$, we have
\begin{equation} 
\xi_{\tilde{d}}
= -3\sum_{j=1}^3
\dfrac{
\left(
\left|\lambda^\prime_{2j1}\right|^2
-\left|\lambda^\prime_{3j1}\right|^2
+\left|\lambda^{\prime}_{21j}\right|^2
-\left|\lambda^\prime_{31j}\right|^2
\right)/M_{\tilde d_{j}}^2}
{g^2/M_W^2} \;.
\label{xi_LQD_degenerate}
\end{equation}


\subsection{Constraints on the R-parity Violating Couplings and Squark/Slepton Masses}

\begin{figure}[t]
	\centering
		\includegraphics[width=8.1cm]{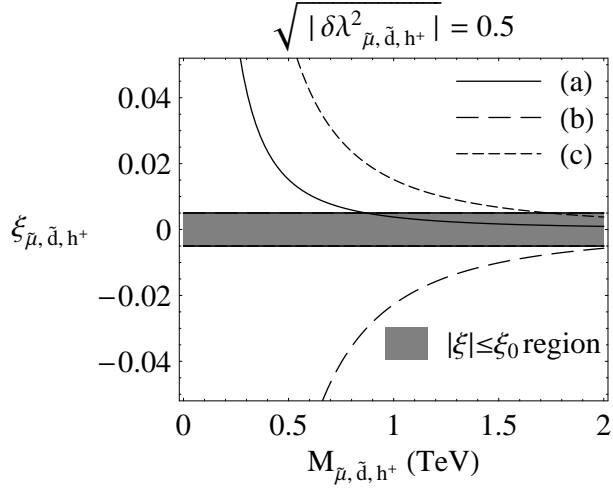}		
	\caption{Dependence of $\xi_{\tilde{\mu},\tilde{d},h}$ 
	 on the smuon, sdown, and $h^\pm$ masses for 
	$\sqrt{\delta\lambda^2_{\tilde{\mu},\tilde{d},h}} = 0.5$ in the 
	(a) $\hat{L}\hat{L}\hat{E}$ R-parity violating interaction;
	(b) $\hat{L}\hat{Q}\hat{D}$ R-parity violating interaction; and 
	(c) the Zee/Babu-Zee models.}
	\label{RparityZeeBabu_xi_mass}
\end{figure}

\begin{figure}[t]
	\centering
		\includegraphics[width=8.1cm]{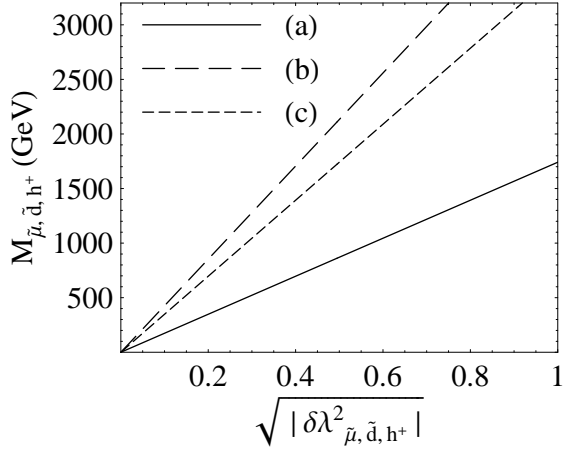}		
	\caption{Lower bounds on
	(a) the smuon mass in the $\hat{L}\hat{L}\hat{E}$ R-parity violating interaction model,	 
	(b) the sdown mass in the $\hat{L}\hat{Q}\hat{D}$ R-parity violating interaction model, and 
	(c) the $h^\pm$ mass in the Zee/Babu-Zee models, respectively. }
	\label{RparityZeeBabu}
\end{figure}

To illustrate our result for R-parity violating interactions,
we simplify the analysis by assuming that only the $\lambda_{122}$ and $\lambda_{132}$ couplings are non-zero for the $\hat{L}\hat{L}\hat{E}$ case, and
only the $\lambda'_{211}$ and $\lambda'_{311}$ couplings are non-zero 
for the $\hat{L}\hat{Q}\hat{D}$ case.
Under these assumptions, only the smuon, $\tilde{e}_2 = \tilde{\mu}$, 
contributes in the first case, and only the sdown, $\tilde{d}_1=\tilde{d}$, 
contributes in the latter.
The corresponding $\xi$'s are 
\begin{eqnarray}
\xi_{\tilde\mu}
& = &+\;\frac{\delta\lambda_{\tilde\mu}^2/M_{\tilde\mu}^2}{(g/M_W)^2}
\;=\;+\frac{1}{4\sqrt{2}G_F} 
\left(\frac{\delta\lambda^2_{\tilde\mu}}{M^2_{\tilde\mu}}\right)
\;,\cr
\xi_{\tilde d}
& = & -6\,\frac{\delta\lambda_{\tilde d}^2/M_{\tilde d}^2}{(g/M_W)^2}
\;=\;-\frac{6}{4\sqrt{2}G_F}
\left(\frac{\delta\lambda^2_{\tilde d}}{M^2_{\tilde d}}\right)
\;,
\end{eqnarray}
where 
\begin{eqnarray}
\delta\lambda^2_{\tilde{\mu}} & \equiv &
|\lambda_{122}|^2 - |\lambda_{132}|^2\;,\cr
\delta\lambda^2_{\tilde{d}} & \equiv & 
|\lambda'_{211}|^2-|\lambda'_{311}|^2\;.
\end{eqnarray}
Fig.~\ref{RparityZeeBabu_xi_mass} shows
how $\xi_{\tilde\mu}$ and $\xi_{\tilde d}$ depend on masses of the 
smuon and the sdown for a specific choice of couplings: 
$\sqrt{\delta\lambda^2_{\tilde\mu}}=\sqrt{\delta\lambda^2_{\tilde d}}=0.5$ 
(we have assumed $\delta\lambda^2_{\tilde d}$ and $\delta\lambda^2_{\tilde\mu}$ to be positive).
The bound $|\xi| \le \xi_0=0.005$ translates into:
\begin{eqnarray}
M_{\tilde\mu} & \ge & \sqrt{|\delta\lambda_{\tilde\mu}^2|}\,\sqrt{\frac{1}{4\sqrt{2}G_F\xi_0}}
\;\approx\;  \sqrt{|\delta\lambda_{\tilde\mu}^2|} \times (1700\,\mathrm{GeV})\;,
\cr
M_{\tilde d} & \ge & \sqrt{|\delta\lambda_{\tilde d}^2|}\,\sqrt{\frac{6}{4\sqrt{2}G_F\xi_0}}
\;\approx\; \sqrt{|\delta\lambda_{\tilde d}^2|} \times (4300\,\mathrm{GeV})\;.
\label{RPV_mass_limits}
\end{eqnarray}
The resulting graphs for the lower mass bounds are shown in 
Fig.~\ref{RparityZeeBabu}.  The regions of the 
$\left(M_{\tilde{\mu}}, \sqrt{|\delta\lambda_{\tilde\mu}^2|}\right)$ 
and 
$\left(M_{\tilde d}, \sqrt{|\delta\lambda_{\tilde d}^2|}\right)$ 
parameter spaces below each of the lines are excluded. 
One can also fix the smuon and sdown masses and obtain upper bounds on the
R-parity violating couplings:
\begin{eqnarray}
\sqrt{|\delta\lambda^2_{\tilde{\mu}}|} & \le & 
\sqrt{4\sqrt{2}G_F\,\xi_0}\;M_{\tilde{\mu}}
\;=\; (0.057)\left(\dfrac{M_{\tilde{\mu}}}{100\,\mathrm{GeV}}\right)\;,\cr
\sqrt{|\delta\lambda^2_{\tilde{d}}|} & \le & 
\sqrt{\dfrac{4\sqrt{2}G_F\,\xi_0}{6}}\; M_{\tilde{d}}
\;=\; (0.023)\left(\dfrac{M_{\tilde{d}}}{100\,\mathrm{GeV}}\right)
\;.
\label{RPV_lambda_limits}
\end{eqnarray}
These relations are actually more useful than Eq.~(\ref{RPV_mass_limits}) 
since if the smuon and sdown exist, their masses will be measured/constrained 
by searches for their pair-production at the LHC, independently of the size of possible R-parity violating couplings.

\begin{table}[ht]
\begin{tabular}{|c||c|c|}
\hline
\ Coupling\ \ & \ Current $2\sigma$ Bound\ \ & \ Observable/Process\ \ \\
\hline\hline
$|\lambda_{122}|$ 
& $0.05\left(\dfrac{M_{\tilde{\mu}_R}}{100\,\mathrm{GeV}}\right)$ 
& \ $V_{ud}$ from nuclear $\beta$ decay/muon decay\ \
\\
\hline
$|\lambda_{132}|$ 
& $0.07\left(\dfrac{M_{\tilde{\mu}_R}}{100\,\mathrm{GeV}}\right)$  
& \ $R_{\tau} = \dfrac{\Gamma(\tau^{-}\rightarrow e^{-} \bar{\nu}_e \nu_\tau)}{\Gamma(\tau^{-}\rightarrow\mu^{-}\bar{\nu}_\mu \nu_\tau)}$\ \ 
\\
\hline
$|\lambda_{122}^{\phantom{*}}\lambda_{132}^*|$
& \ $(2.2\times 10^{-3})
\left(\dfrac{M_{\tilde{\nu}_R}}{100\,\mathrm{GeV}}\right)^2$\ \ 
& $\tau\rightarrow 3\mu$ 
\\
\hline\hline
$|\lambda'_{211}|$ 
& $0.06\left(\dfrac{M_{\tilde{d}_R}}{100\,\mathrm{GeV}}\right)$  
& \ $R_{\pi} = \dfrac{\Gamma(\pi^{-}\rightarrow e^{-} \bar{\nu}_e)}{\Gamma(\pi^{-}\rightarrow\mu^{-}\bar{\nu}_\mu)}$\ \ 
\\
\hline
$|\lambda'_{311}|$ 
& $0.12\left(\dfrac{M_{\tilde{d}_R}}{100\,\mathrm{GeV}}\right)$  
& \ $R_{\tau\pi} = \dfrac{\Gamma(\tau^{-}\rightarrow\pi^{-} \nu_\tau)}{\Gamma(\pi^{-}\rightarrow\mu^{-}\nu_\mu)}$\ \ 
\\
\hline
\end{tabular}
\caption{Current $2\sigma$ bounds on R-parity violating couplings from Ref.~\cite{Barbier:2004ez}. These bounds assume that each coupling is non-zero
only one at a time.}
\label{CurrentRparityBounds}
\end{table}


Current bounds on R-parity violating couplings come from a variety of
sources \cite{Barbier:2004ez,Rparity_limits}.
The current indirect bounds of the four couplings under consideration
from low-energy experiments are listed in Table~\ref{CurrentRparityBounds}.
Comparison with Eq.~(\ref{RPV_lambda_limits}) shows that the bounds on
$\lambda_{122}$ and $\lambda_{132}$ are already fairly tight, and
neutrino oscillation will do little to improve them.
On the other hand, the bounds on $\lambda'_{211}$ and $\lambda'_{311}$
can potentially be improved by factors of roughly $2.5$ and $5$, respectively.

Bounds on R-parity violating couplings from $ep$ and $p\bar{p}$ 
colliders come from searches for $s$-channel resonant production of sparticles.
The bounds from the $ep$ collider HERA necessarily involve the
couplings $\lambda'_{1jk}$ since the squark must couple to
the first generation lepton (electron or positron)
\cite{Chekanov:2003af,Chekanov:2005au,Adloff:2003jm,Aktas:2004ij}
so we will not discuss them here.
The bound from the Tevatron comes from the analysis of D0 which 
looked for the R-parity violating processes 
$d\bar{u}\rightarrow \tilde{\mu}$ or $d\bar{d}\rightarrow\tilde{\nu}_\mu$,
which occur if $\lambda'_{211}\neq 0$, 
followed by the decay of the slepton via the R-parity conserving processes
$\tilde{\mu}\rightarrow \tilde{\chi}^0_{1,2,3,4}\,\mu$ or
$\tilde{\nu}_\mu\rightarrow \tilde{\chi}^{\pm}_{1,2}\,\mu$ \cite{Abazov:2006ii}.
The neutralinos and charginos produced in these processes cascade decay
down to the $\tilde{\chi}^0_1$ 
(the assumed lightest supersymmetric particle, or LSP)
which decays via a virtual smuon, muon-sneutrino, or squark though 
the R-parity violating $\lambda'_{211}$ coupling again into a muon and two jets, 
giving 2 muons in the final state.
The bound on the value of $\lambda'_{211}$ from 
this analysis depends in a complicated manner on all the masses of the particles
involved in the processes.  If one uses a 
minimal supergravity (mSUGRA) framework \cite{Martin:1997ns}
with $\tan\beta=5$, $\mu<0$, and $A_0=0$, then the 95\% bound 
is $\lambda'_{211}\le 0.1$ 
assuming $M_{\tilde{\mu}} = 363\,\mathrm{GeV}$ \cite{Abazov:2006ii}.
A similar bound would result from Eq.~(\ref{RPV_lambda_limits}) if $M_{\tilde{d}} = 460\,\mathrm{GeV}$.
However, since squarks are generically much heavier than sleptons 
\cite{Martin:1997ns},
the existing D0 bound is effectively stronger than the
potential bound from $|\xi|\le\xi_0$.


\section{Extended Higgs Models}


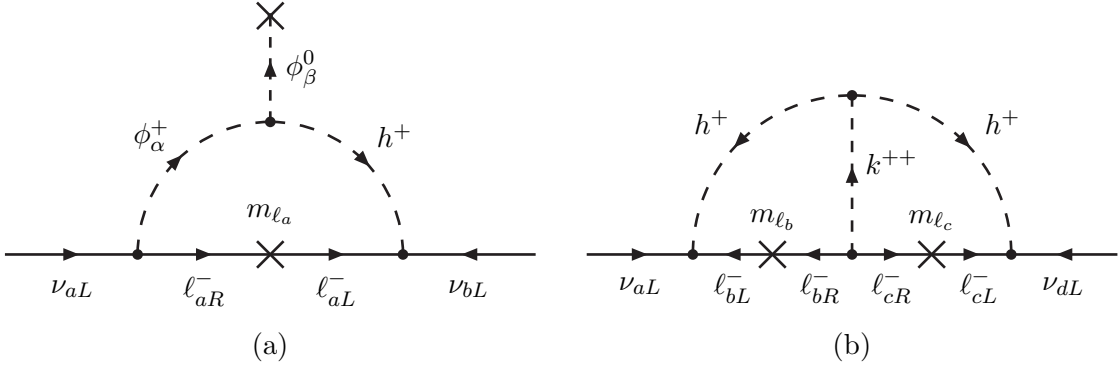
\begin{figure}[ht]
\centering
    \begin{picture}(440,140)(-100,-60) 
    \SetScale{1}  
    \SetColor{Black}
    
    \SetWidth{1}
    \ArrowLine(-100,-25)(-50,-25)
    \ArrowLine(-50,-25)(0,-25)
    \ArrowLine(0,-25)(50,-25)
    \ArrowLine(100,-25)(50,-25)
    \DashArrowLine(0,25)(0,65){5}
    \DashArrowArcn(0,-25)(50,90,0){5}
    \DashArrowArcn(0,-25)(50,180,90){5}
    \Line(-5,60)(5,70)       
    \Line(5,60)(-5,70)
    	\Vertex(0,65){1}
    \Vertex(0,25){2}
    \Vertex(-50,-25){2}
    \Vertex(50,-25){2}
    \Line(-5,-30)(5,-20)       
    \Line(5,-30)(-5,-20)
    	\Text(0,-10)[]{$m_{\ell_a}$}       
    \Text(-75,-38)[]{$\nu_{aL}$}
    \Text(75,-38)[]{$\nu_{bL}$}
	\Text(-25,-38)[]{$\ell_{aR}^-$}
	\Text(25,-38)[]{$\ell_{aL}^-$}
    \Text(12,45)[]{$\phi_\beta^{0}$}
	\Text(47,20)[]{$h^+$}
	\Text(-45,20)[]{$\phi_\alpha^+$}
	\Text(0,-60)[]{(a)}
    
    \SetOffset(220,0)
    \SetWidth{1}
    \ArrowLine(-100,-25)(-60,-25)
    \ArrowLine(-30,-25)(-60,-25)
    \ArrowLine(0,-25)(-30,-25)
    \ArrowLine(100,-25)(60,-25)
    \ArrowLine(30,-25)(60,-25)
    \ArrowLine(0,-25)(30,-25)
    \DashArrowLine(0,-25)(0,35){5}
    \DashArrowArcn(0,-25)(60,90,0){5}
    \DashArrowArc(0,-25)(60,90,180){5}
    \Line(-35,-30)(-25,-20)       
    \Line(-25,-30)(-35,-20)
    \Line(35,-30)(25,-20)       
    \Line(25,-30)(35,-20)
    \Vertex(0,35){2}
    \Vertex(0,-25){2}       
    \Vertex(-60,-25){2}
    \Vertex(60,-25){2}       
    \Text(-80,-37)[]{$\nu_{aL}$}
    \Text(80,-37)[]{$\nu_{dL}$}
    	\Text(-30,-12)[]{$m_{\ell_b}$}       
    	\Text(30,-12)[]{$m_{\ell_c}$}       
	\Text(-45,-37)[]{$\ell_{bL}^-$}
	\Text(48,-37)[]{$\ell_{cL}^-$}
	\Text(-12,-37)[]{$\ell_{bR}^-$}
	\Text(15,-37)[]{$\ell_{cR}^-$}
    \Text(15,10)[]{$k^{++}$}
	\Text(57,25)[]{$h^+$}
	\Text(-53,25)[]{$h^+$}
	\Text(0,-60)[]{(b)}

    \end{picture}
	\caption{Diagrams which generate the Majorana masses and mixings of the neutrino in the (a) Zee \cite{Zee:1980ai} and (b) Babu-Zee \cite{Babu:1988ki} models.
	}
	\label{ZeeBabuNuMass}
\end{figure}    

Most models, including the Standard Model (SM) and its various
extensions, possess Higgs sectors which
distinguish among the different generation fermions.
The models discussed in section~\ref{ZprimeSection}
are necessarily so, and so are the Zee \cite{Zee:1980ai} and 
Babu-Zee \cite{Babu:1988ki} models of neutrino mass, 
as well as various triplet Higgs models \cite{triplet_higgs}.
As representative cases, we consider the effect of 
the singlet Higgs in the Zee and Babu-Zee models, and
that of a triplet Higgs with hypercharge $Y=+1$ ($Q_{em}=I_3+Y$).

\subsection{Singlet Higgs in the Zee and Babu-Zee Models}

In the Zee \cite{Zee:1980ai} and Babu-Zee \cite{Babu:1988ki} models,
an isosinglet scalar $h^+$ with hypercharge $Y=+1$ is introduced,
which couples to left-handed lepton doublets as
\begin{equation} \label{L_Babu}
\mathcal{L}_h
\;=\; \lambda_{ab}
\left(\,\ell_{aL}^\mathrm{T}C\,i\sigma_2\,\ell_{bL}^{\phantom{\mathrm{T}}}\,\right) h^+ + h.c.
\;=\; \lambda_{ab} 
\left(\,\overline{\ell^c_{aL}}\,i\sigma_2\,\ell_{bL}^{\phantom{\mathrm{T}}}\,\right) h^+ + h.c.\;,
\end{equation}
where $(ab)$ are flavor indices: $a,b=e,\mu,\tau$. 
The hypercharge assignment prohibits the $h^\pm$ fields from having a similar interaction with the quarks. 
Due to $SU(2)$ gauge invariance,
the couplings $\lambda_{ab}$ are antisymmetric: $\lambda_{ab}=-\lambda_{ba}$.
This interaction is analogous to the R-parity violating $\hat{L}\hat{L}\hat{E}$ 
coupling with $h^\pm$ playing the role of the slepton.
 
In the Zee model \cite{Zee:1980ai}, in addition to the $h^\pm$,
two or more $SU(2)$ doublets $\phi_\alpha$ ($\alpha=1,2,\cdots$) 
with hypercharge $Y=-\frac{1}{2}$
are introduced which couple to the $h^\pm$ via
\begin{equation}
\mathcal{L}_{\phi\phi h} \;=\;
M_{\alpha\beta}\left(\phi^\mathrm{T}_\alpha\,i\tau_2\,\phi_\beta^{\phantom{\mathrm{T}}}\right) h^+ 
+ h.c.\;,
\end{equation}
and to the fermions in the usual fashion.
The couplings $M_{\alpha\beta}$ are antisymmetric, just like $\lambda_{ab}$,
which necessitates the introduction of more than one doublet. 
In this model, Majorana masses and mixings of the neutrinos are generated at one-loop as shown in Fig.~\ref{ZeeBabuNuMass}a.
The extra doublets can also contribute to neutrino oscillation
depending on their Yukawa couplings to the leptons, but we will assume 
that their effect is negligible compared to that of the $h^\pm$.

In the Babu-Zee model \cite{Babu:1988ki}, in addition to the $h^\pm$,
another isosinglet scalar $k^{++}$ with hypercharge $Y=+2$
is introduced which couples to the right-handed leptons and $h^\pm$ via
\begin{equation}
\mathcal{L}_k \;=\; \lambda'_{ab}
\left(\,\overline{e_{aR}^c}\,e_{bR}^{\phantom{c}}\,\right) k^{++} - M\,h^{+}h^{+}k^{--}
+ h.c.\;,
\end{equation}
where $\lambda'_{ab}=\lambda'_{ba}$.
In this model, Majorana masses and mixings of the neutrinos are generated at the two-loop level as shown in Fig.~\ref{ZeeBabuNuMass}b. 
In this case, the extra scalar, $k$, does not contribute to
neutrino oscillation.

\begin{figure}[t]
\centering
    \begin{picture}(400,100)(-100,-60) 
    \SetWidth{1}
    \SetScale{1}  
    \SetColor{Black}
    
    \ArrowLine(-80,-40)(-40,0)
    \ArrowLine(-80,40)(-40,0)
    \DashArrowLine(-40,0)(40,0){5}
    \ArrowLine(40,0)(80,40)
    \ArrowLine(40,0)(80,-40)
    \Vertex(-40,0){2}
    \Vertex(40,0){2}       
    \Text(-70,50)[]{$\nu_\ell(k)$}
    \Text(70,50)[]{$e(p)$}
    \Text(-70,-50)[]{$e(p)$}
    \Text(70,-50)[]{$\nu_\ell(k)$}
    \Text(0,15)[]{$h^-$}
    \Text(0,-20)[]{$p+k$}
    \Text(-70,0)[]{$+i\, \lambda_{e\ell}$}
    \Text(70,0)[]{$+i\, \lambda_{e\ell}$}
    
    \SetWidth{0.5}
    \LongArrow(-15,-10)(15,-10)

	\SetOffset(200,0)
	\SetWidth{1}
    \ArrowLine(-80,-40)(-40,0)
    \ArrowLine(-80,40)(-40,0)
    \DashArrowLine(-40,0)(40,0){5}
    \ArrowLine(40,0)(80,40)
    \ArrowLine(40,0)(80,-40)
    \Vertex(-40,0){2}
    \Vertex(40,0){2}       
    \Text(-70,50)[]{$\nu_\ell(k)$}
    \Text(70,50)[]{$e(p)$}
    \Text(-70,-50)[]{$e(p)$}
    \Text(70,-50)[]{$\nu_\ell(k)$}
    \Text(0,15)[]{$\Delta^-$}
    \Text(0,-20)[]{$p+k$}
    \Text(-70,0)[]{$+i\, \lambda'_{e\ell}$}
    \Text(70,0)[]{$+i\, \lambda'_{e\ell}$}
    
    \SetWidth{0.5}
    \LongArrow(-15,-10)(15,-10)

    \end{picture}
	\caption{
	Contribution to neutrino oscillation matter effects from a singly-charged Higgs in the Zee, Babu-Zee, and $Y=1$ Triplet Higgs models.
	}
	\label{ExHiggs}
\end{figure}
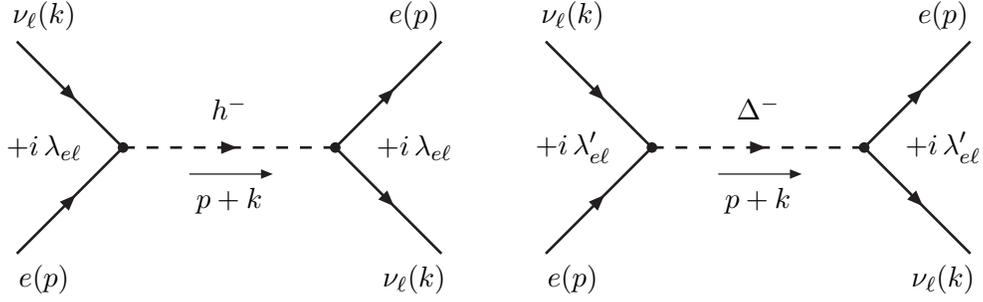

Expanding Eq.~(\ref{L_Babu}), we obtain 
\begin{equation} \label{L_Babu-2}
\mathcal{L}\;=\;2 
\left[\, \lambda_{e\mu}
\left(\,\overline{\nu^c_{e L}}\mu_L^{\phantom{c}}-\overline{\nu^c_{\mu L}}e_L^{\phantom{c}}\,\right)
+\lambda_{e\tau}
\left(\,\overline{\nu^c_{e L}}\tau_L^{\phantom{c}}-\overline{\nu^c_{\tau L}}e_L^{\phantom{c}}\,\right)
+\lambda_{\mu\tau}
\left(\,\overline{\nu^c_{\mu L}}\tau_L^{\phantom{c}}-\overline{\nu^c_{\tau L}}\mu_L^{\phantom{c}}\,\right)
\,\right] h^+ +h.c.
\end{equation}
Keeping only the terms that are relevant for neutrino oscillation matter effects, we have
\begin{equation}
-2
\left(\,\lambda_{e\mu}\,\overline{\nu^c_{\mu L}}\,e_L^{\phantom{c}}
+ \lambda_{e\tau}\,\overline{\nu^c_{\tau L}}\,e_L^{\phantom{c}}
\,\right) h^+ +h.c.
\label{ZeeBabuRelevant}
\end{equation}
The corresponding Feynman diagram is shown in Fig.~\ref{ExHiggs}. 

Calculations similar to those for the $S_1$ leptoquark yield
\begin{equation}
V_{\nu_{\mu}}\;=\;-N\,\dfrac{|\lambda_{e\mu}|^2}{M^2_{h}}\;,\qquad
V_{\nu_{\tau}}\;=\;-N\,\dfrac{|\lambda_{e\tau}|^2}{M^2_{h}}\;,
\end{equation}
and
\begin{equation} \label{xi_babu}
\xi_{h}\;=\;
\dfrac{V_{\nu_{\tau}}-V_{\nu_{\mu}}}{V_{CC}}
\;=\;4\,\dfrac{(|\lambda_{e\mu}|^2-|\lambda_{e\tau}|^2)/M^2_{h}}{(g/M_W)^2}
\;=\;+\frac{1}{\sqrt{2}G_F}
\left( \frac{\delta\lambda^2_{h}}{M^2_{h}} \right)
\;,
\end{equation}
where we have defined 
$\delta \lambda^2_h \equiv |\lambda_{e\mu}|^2-|\lambda_{e\tau}|^2$. 
The dependence of $\xi_{h}$ on the $h^\pm$ mass is plotted in 
Fig.~\ref{RparityZeeBabu_xi_mass} for the case $\sqrt{\delta\lambda^2_{h}}=0.5$,
where we have assumed $\delta\lambda^2_{h}>0$.
The bound $|\xi|\le\xi_0=0.005$ translates into
\begin{equation}
\left|\dfrac{\delta\lambda^2_{h}}{M^2_{h}}\right|
\;\le\; \sqrt{2}G_F\,\xi_0
\;=\; (8.2\times 10^{-8})\,\mathrm{GeV}^{-2}\;,
\label{ExHiggsConstraint}
\end{equation}
or
\begin{equation}
M_{h}\;\ge\;
\sqrt{\frac{|\delta\lambda_{h}^2|}{\sqrt{2} G_F\, \xi_0}}
\;\approx\; 
\sqrt{ |\delta\lambda_{h}^2| }\times(3500\,\mathrm{GeV}) \;.
\end{equation}
This result is represented graphically in Fig.~\ref{RparityZeeBabu}. 
The region of the $(M_{h}, \sqrt{|\delta\lambda_{h}^2|})$ 
parameter space below the constructed line would be excluded. 

A constraint on the exact same combination of the couplings and mass
of the $h^\pm$ as above exists from $\tau$ decay data:
The measured value of the
$\tau^{-}\rightarrow \nu_\tau e^-\bar{\nu}_e$ branching fraction imposes
the constraint \cite{Babu:2002uu}
\begin{equation}
\left|\dfrac{\delta\lambda^2_{h}}{M^2_{h}}\right|
\;\le\; (3.4\times 10^{-8})\,\mathrm{GeV}^{-2}\;,
\end{equation}
which is clearly stronger than Eq.~(\ref{ExHiggsConstraint}).

\subsection{Triplet Higgs with $Y=+1$}

We denote the components of an isotriplet Higgs with hypercharge $Y=+1$ as
\begin{equation}
\left[ \begin{array}{l} \Delta^{++} \\ \Delta^{+} \\ \Delta^{0} \end{array}
\right]\;.
\end{equation}
It is customary to write this in $2\times 2$ matrix form:
\begin{equation}
\Delta \;\equiv\;
\dfrac{1}{\sqrt{2}}
\left[
  \Delta^{0} \left(\dfrac{\sigma_1-i\sigma_2}{\sqrt{2}}\right)
+ \Delta^+   \sigma_3 
+ \Delta^{++}\left(\dfrac{\sigma_1+i\sigma_2}{\sqrt{2}}\right)
\right]
\;=\;
\left[ \begin{array}{cc}
\Delta^{+}/\sqrt{2} & \Delta^{++} \\
\Delta^{0} & -\Delta^{+}/\sqrt{2}
\end{array} \right] \;.
\end{equation}
The coupling of $\Delta$ to the leptons is then
\begin{equation}
\mathcal{L}_{\Delta}
\;=\; \sqrt{2}\lambda'_{ab}
\left( \ell_{aL}^{\mathrm{T}} C\,i\sigma_2\,\Delta\, \ell_{bL}^{\phantom{\mathrm{T}}} \right) + h.c. 
\;=\; \sqrt{2}\lambda'_{ab}
\left( \overline{\ell_{aL}^c}\,i\sigma_2\,\Delta\, \ell_{bL}^{\phantom{\mathrm{T}}} \right) + h.c.
\label{L_TripletHiggs}
\end{equation}
This time, the couplings are symmetric in the flavor indices $\lambda'_{ab}=\lambda'_{ba}$, and the factor of $\sqrt{2}$ is thrown in for latter convenience.  Expanding out, we find
\begin{equation}
\mathcal{L}_{\Delta}
\;=\;\lambda'_{ab}
\left[\, \sqrt{2} 
\left(\overline{\nu_{aL}^c}\nu_{bL}^{\phantom{c}}\right)\Delta^0
-\left(\overline{\nu_{aL}^c}e_{bL}^{\phantom{c}}+\overline{e_{aL}^c}\nu_{bL}^{\phantom{c}}\right)\Delta^+
-\sqrt{2}\left(\overline{e_{aL}^c}e_{bL}^{\phantom{c}}\right) \Delta^{++}
\,\right] + h.c. 
\end{equation}
and the terms relevant to neutrino oscillation in matter are:
\begin{equation}
-2
\left(\,
\lambda'_{ee}\,\overline{\nu^c_{e L}}\,e_L^{\phantom{c}}
+ \lambda'_{e\mu}\,\overline{\nu^c_{\mu L}}\,e_L^{\phantom{c}}
+ \lambda'_{e\tau}\,\overline{\nu^c_{\tau L}}\,e_L^{\phantom{c}}
\,\right) \Delta^+ +h.c.
\end{equation}
Of these, the $\lambda'_{ee}$ term does not affect $\xi$, while
the other terms are precisely the same as those listed in Eq.~(\ref{ZeeBabuRelevant}).
So without further calculations, we can conclude that all the
results of the previous subsection apply in this case also.

\begin{table}[t]
\begin{tabular}{|l||c|c|}
\hline
Model & \ Stronger than existing bounds?\ \ & \ Competitive with LHC?\ \ \\
\hline\hline
Gauged $L_e-L_\mu$ and $L_e-L_\tau$ 
& No
& ---
\\
\hline
Gauged $B-3L_\tau$
& Yes
& Yes
\\
\hline
Topcolor Assisted Technicolor
& No
& ---
\\
\hline
Leptoquarks
& Yes
& Yes${}^*$
\\
\hline
R-parity violation
& No
& ---
\\
\hline
Zee, Babu-Zee, Triplet Higgs
& No
& ---
\\
\hline
\end{tabular}
\caption{The result of our survey. The potential bound from
$|\xi|\le\xi_0=0.005$ is compared with existing bounds, and the expected
bounds from the LHC.  If the existing bound is already stronger, 
no comparison with the LHC bound is made.
${}^*$The leptoquark bound will be competitive with the LHC, 
provided that $\sqrt{|C_{LQ}||\delta\lambda^2_{LQ}|}= O(1)$.}
\label{Comparison}
\end{table}

\section{Summary and Conclusions}

In this paper, we surveyed the potential constraints 
on various models of new physics which could be obtained from
a hypothetical Fermilab$\rightarrow$HyperKamiokande, or
similar type of experiment.
We assumed that the parameter $\xi$, defined in Eq.~(\ref{xi-def}),
could be constrained to $|\xi|\le\xi_0 = 0.005$ at the 99\% confidence
level. This places a constraint on the couplings and masses of 
new particles that are exchanged between the neutrinos and matter
fermions.

Table~\ref{Comparison} summarizes our result.
Of the models surveyed, the potential bound on gauged $B-3L_\tau$
from $|\xi|\le\xi_0$ can be expected to be stronger than the
expected bound from the LHC.  Bounds on generation non-diagonal 
leptoquarks can be competitive if 
$\sqrt{|C_{LQ}||\delta\lambda^2_{LQ}|}=O(1)$.
For these cases, neutrino oscillation can be used as an independent
check in the event that such new physics is discovered at the LHC.

All the other models are already well constrained by existing
experiments, either indirectly by low-energy precision measurements, or
by direct searches at colliders.  Generically, the couplings and masses of 
new particles that couple only to leptons are well constrained by lepton
universality, 
while their contribution to neutrino oscillation tend to be suppressed since they only interact with the electrons in matter.
This tends to render the existing bound stronger than the potential bound from
$|\xi|\ge \xi_0$.

Topcolor assisted technicolor, and R-parity violating $LQD$ couplings
involve interactions with the quarks in matter, but they too belong to the list of already well-constrained models.
For the $Z'$ in topcolor assisted technicolor, the proton and electron contributions to neutrino oscillation cancel, 
just as for the Standard Model $Z$, and the coupling
is also fixed to a small value, which results in a weak bound from
$|\xi|\le\xi_0$.
For the $LQD$ coupling, restriction to minimal supergravity provided an
extra constraint which strengthened the existing bound.

The fact that only a limited number of models (at least among those
we surveyed) can be well constrained by $|\xi|\le\xi_0$ means, conversely, 
that if a non-zero $\xi$ is observed in neutrino oscillation, the list
of possible new physics that could lead to such an effect is
also limited.
This could, in principle, help distinguish among possible new physics
which have the same type of signature (\textit{e.g.} a leptoquark which may, or may not be generation diagonal) at the LHC.

\section*{Acknowledgments}

We would like to thank Andrew Akeroyd, Mayumi Aoki, Masafumi Kurachi, 
Sarira Sahu, Hiroaki Sugiyama, Nguyen Phuoc Xuan, and Marek Zralek for helpful discussions.
Takeuchi would like to thank the particle theory group at
Ochanomizu University for their hospitality during the summer of 2006.
Portions of this work have been presented at invited talks at
the 19th and 20th Workshops on Cosmic Neutrinos at the ICRR, Kashiwa (Takeuchi 7/6/06, Okamura 2/20/07); 
NuFact2006 at U.C. Irvine (Okamura 8/29/06);
and the Joint Meeting of Pacific Region Particle Physics Communities in Hawaii (Honda 10/31/06).
We thank the conveners of these meetings for providing us with so many opportunities to present our work.
This research was supported in part by the U.S. Department of Energy, 
grant DE--FG05--92ER40709, Task A (Kao, Pronin, and Takeuchi).


\end{document}